\documentstyle[psfig]{mn}

\newif\ifAMStwofonts
\AMStwofontstrue

\newcommand{\be}{\begin{equation}}
\newcommand{\ee}{\end{equation}}
\newcommand{\ba}{\begin{eqnarray}}
\newcommand{\ea}{\end{eqnarray}}
\newcommand{\brr}{\begin{array}}
\newcommand{\err}{\end{array}}
\newcommand{\bc}{\begin{center}}
\newcommand{\ec}{\end{center}}
\newcommand{\hm}{\,h^{-1}{\rm Mpc}}
\newcommand{\hk}{\,h^{-1}{\rm kpc}}

\newcommand{\vel}{\,{\rm km\,s^{-1}}}

\newcommand{\mincir}{\raise
  -2.truept\hbox{\rlap{\hbox{$\sim$}}\raise5.truept \hbox{$<$}\ }}
\newcommand{\magcir}{\raise
  -2.truept\hbox{\rlap{\hbox{$\sim$}}\raise5.truept \hbox{$>$}\ }}
\newcommand{\siml}{\raise
  -2.truept\hbox{\rlap{\hbox{$\sim$}}\raise5.truept \hbox{$<$}\ }}
\newcommand{\simg}{\raise
  -2.truept\hbox{\rlap{\hbox{$\sim$}}\raise5.truept \hbox{$>$}\ }}

\ifoldfss
\ifCUPmtlplainloaded \else
\NewTextAlphabet{textbfit} {cmbxti10} {}
\NewTextAlphabet{textbfss} {cmssbx10} {}
\NewMathAlphabet{mathbfit} {cmbxti10} {} 
\NewMathAlphabet{mathbfss} {cmssbx10} {} 
\fi
\ifAMStwofonts
\ifCUPmtlplainloaded \else
\NewSymbolFont{upmath} {eurm10}
\NewSymbolFont{AMSa} {msam10}
\NewMathSymbol{\upi}     {0}{upmath}{19}
\NewMathSymbol{\umu}     {0}{upmath}{16}
\NewMathSymbol{\upartial}{0}{upmath}{40}
\NewMathSymbol{\leqslant}{3}{AMSa}{36}
\NewMathSymbol{\geqslant}{3}{AMSa}{3E}

\fi
\fi
\fi 

\ifnfssone
\newmathalphabet{\mathit}
\addtoversion{normal}{\mathit}{cmr}{m}{it}
\addtoversion{bold}{\mathit}{cmr}{bx}{it}
\newmathalphabet{\mathbfit} 
\addtoversion{normal}{\mathbfit}{cmr}{bx}{it}
\addtoversion{bold}{\mathbfit}{cmr}{bx}{it}
\newmathalphabet{\mathbfss} 
\addtoversion{normal}{\mathbfss}{cmss}{bx}{n}
\addtoversion{bold}{\mathbfss}{cmss}{bx}{n}
\ifAMStwofonts
\ifCUPmtlplainloaded \else
\UseAMStwoboldmath
\makeatletter
\new@mathgroup\upmath@group
\define@mathgroup\mv@normal\upmath@group{eur}{m}{n}
\define@mathgroup\mv@bold\upmath@group{eur}{b}{n}
\edef\UPM{\hexnumber\upmath@group}
\new@mathgroup\amsa@group
\define@mathgroup\mv@normal\amsa@group{msa}{m}{n}
\define@mathgroup\mv@bold\amsa@group{msa}{m}{n}
\edef\AMSa{\hexnumber\amsa@group}
\makeatother
\mathchardef\upi="0\UPM19
\mathchardef\umu="0\UPM16
\mathchardef\upartial="0\UPM40
\mathchardef\leqslant="3\AMSa36
\mathchardef\geqslant="3\AMSa3E
\fi
\fi
\fi 

\ifnfsstwo
\DeclareMathAlphabet{\mathbfit}{OT1}{cmr}{bx}{it}
\SetMathAlphabet\mathbfit{bold}{OT1}{cmr}{bx}{it}
\DeclareMathAlphabet{\mathbfss}{OT1}{cmss}{bx}{n}
\SetMathAlphabet\mathbfss{bold}{OT1}{cmss}{bx}{n}
\ifAMStwofonts
\ifCUPmtlplainloaded \else
\DeclareSymbolFont{UPM}{U}{eur}{m}{n}
\SetSymbolFont{UPM}{bold}{U}{eur}{b}{n}
\DeclareSymbolFont{AMSa}{U}{msa}{m}{n}
\DeclareMathSymbol{\upi}{0}{UPM}{"19}
\DeclareMathSymbol{\umu}{0}{UPM}{"16}
\DeclareMathSymbol{\upartial}{0}{UPM}{"40}
\DeclareMathSymbol{\leqslant}{3}{AMSa}{"36}
\DeclareMathSymbol{\geqslant}{3}{AMSa}{"3E}
\fi
\fi
\fi 

\ifCUPmtlplainloaded \else
\ifAMStwofonts \else 
\def\upi{\pi}
\def\umu{\mu}
\def\upartial{\partial}
\fi
\fi

\title[Entropy amplification
in simulated groups and clusters] {Entropy amplification
from energy feedback in simulated galaxy groups and clusters}
  \author[S. Borgani et al.]  {S. Borgani$^{1,2}$, 
  A. Finoguenov$^3$, S.T. Kay $^{4,5}$, T.J. Ponman$^6$, V. Springel$^7$, \\~ \\
\LARGE{P. Tozzi$^8$, G.M. Voit$^9$} \\ ~\\ 
$^1$ Dipartimento di Astronomia dell'Universit\`a di Trieste, via
  Tiepolo 11, 34131 Trieste, Italy (borgani@ts.astro.it)\\
$^2$ INFN -- National Institute for Nuclear Physics, Trieste,
  Italy\\ 
$^3$ Max-Planck-Institut f\"ur extraterrestrische Physik,
  Giessenbachstrasse, 85748 Garching, Germany (alexis@xray.mpe.mpg.de)\\
$^4$ Astronomy Centre, Department of Physics and Astronomy, University
  of Sussex, Brighton BN1 9QH, UK \\
$^5$ Astrophysics, Keble Road, Oxford OX1 3RH, UK (skay@astro.ox.ac.uk)\\
$^6$ School of Physics and Astronomy, University of Birmingham,
  Edgbaston, Birmingham B15 2TT, UK (tjp@star.sr.bham.ac.uk)\\
$^7$ Max-Planck-Institut f\"ur Astrophysik, Karl-Schwarzschild Strasse
  1, Garching bei M\"unchen, Germany (volker@mpa-garching.mpg.de)\\
$^8$ INAF -- Osservatorio Astronomico di Trieste, via Tiepolo 11,
  34131 Trieste, Italy (tozzi@ts.astro.it)\\
$^9$ Department of Physics and Astronomy, Michigan State University,
  BPS Building, East Lansing, MI 48824, USA (voit@pa.msu.edu) 
}

\pubyear{2003}

\begin{document}
\label{firstpage}
\maketitle

\begin{abstract}
  We use hydrodynamical simulations of galaxy clusters and groups to
  study the effect of pre--heating on the entropy structure of the
  intra--cluster medium (ICM). Our simulations account for
  non--gravitational heating of the gas either by imposing a minimum
  entropy floor at redshift $z_h=3$ in adiabatic simulations, or by
  considering feedback by galactic winds powered by supernova (SN)
  energy in runs that include radiative cooling and star formation.
  In the adiabatic simulations we find that the entropy is increased
  out to the external regions of the simulated halos as a consequence
  of the transition from clumpy to smooth accretion induced by extra
  heating. This result is in line with the predictions of the
  semi--analytical model by Voit et al. However, the introduction of
  radiative cooling substantially reduces this entropy amplification
  effect.  While we find that galactic winds of increasing strength
  are effective in regulating star formation, they have a negligible
  effect on the entropy profile of cluster--sized halos.  Only in
  models where the action of the winds is complemented with diffuse
  heating corresponding to a pre--collapse entropy do we find a
  sizable entropy amplification out to the virial radius of the
  groups. Observational evidence for entropy amplification in the
  outskirts of galaxy clusters and groups therefore favours a scenario
  for feedback that distributes heating energy in a more diffuse way
  than predicted by the model for galactic winds from SN explosions
  explored here.
\end{abstract}


\section{Introduction}
The thermodynamic properties of the intra--cluster medium (ICM) reflect the
full history of the cosmic evolution of baryons. This evolution is primarily
driven by gravity, which shapes the gas distribution on large scales and
determines the hierarchical assembly of cosmic structures, but also by a
complex interplay with the hydrodynamic processes of star formation and galaxy
evolution (see Rosati et al. 2002; Voit 2005 for recent reviews, and
references therein). Due to the high density and temperature reached by 
gas in the environment of galaxy clusters and groups, X--ray observations
provided so far the most powerful diagnostics on its thermal properties. 

The simplest model for the evolution of the ICM accounts only for the
action of gravity (Kaiser 1986). This picture is said to be
self--similar because it predicts that galaxy groups and clusters of
different richness are essentially scaled versions of each
other. Based on this realization, one can then derive unique scaling
relations between X--ray observables.  For instance, the X--ray
luminosity, $L_X$, should scale with the gas temperature $T$ as
$L_X\propto T^2$; the ``entropy'' $S=T/n_e^{2/3}$ ($n_e$ is the
electron number density) should scale linearly with temperature owing
to the self--similarity of the gas density profiles. Furthermore,
spherical gas accretion in a NFW halo (Navarro, Frenk \& White 1997)
predicts entropy to scale with the cluster--centric distance as
$S\propto R^{1.1}$ (Tozzi \& Norman 2001).

While adiabatic hydrodynamical simulation of clusters have validated
this simple picture (e.g., Navarro et al. 1995; Eke et al. 1998;
Borgani et al.  2001), a number of observations have now
established that gravity cannot be the only player: the gas density
profiles in the central regions of groups and poor clusters are
observed to be shallower than in the self-similar model, and the
relative entropy level is correspondingly higher than in rich clusters
(e.g., Ponman et al. 1999; Finoguenov et al. 2002; Ponman et al.
2003). As a consequence, smaller systems are relatively underluminous,
thus producing too steep an $L_X$--$T$ relation, $L_X\propto T^{\sim
3}$ (e.g., Markevitch 1998; Arnaud \& Evrard 1999; Sanderson et
al. 2003).

These observational facts call for additional physics in the form of
non--gravitational processes, capable of breaking the self--similarity of the
ICM properties. As a solution, a number of authors have proposed the presence
of extra heating, possibly occurring before the cluster collapse (e.g. Evrard
\& Henry 1991; Kaiser 1991; Bower 1997; Cavaliere, Menci \& Tozzi 1998; Balogh
et al. 1999; Bialek, Evrard \& Mohr 2001; Tozzi \& Norman 2001; Borgani et al.
2002; Dos Santos \& Dor\'e 2002), or of radiative cooling (e.g., Bryan 2000;
Voit \& Bryan 2001; Wu \& Xue 2002), or, perhaps more likely, of a combination
of these two effects (e.g., Voit et al. 2002; Muanwong et al.  2002; Tornatore
et al. 2003; Borgani et al. 2004; Kay et al. 2003, 2004). Radiative cooling
determines the minimum entropy level of the X--ray emitting gas through a
selective removal of gas with short cooling time. On the other hand,
non--gravitational heating affects the amount of gas which can cool down to
form stars, thereby preventing a cooling runaway. Independent of the details
of the model and the nature of the assumed heating energy, it has been common
prejudice that the ICM entropy would only be affected in the central regions
of clusters, while the self--similarity would be preserved in the halo
outskirts, where dynamics is still dominated by gravity.

However, based on a semi--analytic model for gas accretion in clusters, Voit
et al. (2003) have recently predicted that a smoothing of the gas density due
to preheating in infalling sub--halos would boost the entropy production at
the accretion shock of clusters. As a result, an excess of entropy with
respect to the prediction of the self--similar model is generated in the
cluster outskirts.  This effect due to smooth accretion should be more
important for poorer systems, since their accretion is dominated by smaller
clumps which are more affected by pre--heating. Therefore, entropy generation
is expected to be more efficient in smaller systems, thus boosting their
entropy profiles more than those of more massive structures. 

Observational support for this model has been provided by Ponman et
al.  (2003). Using ROSAT/ASCA data for a large set of clusters and
groups, they found that the entropy level in the outskirts is higher
than predicted by self--similar scaling, in fact by a larger amount
for poorer systems.  Ponman et al.  (2003) also gave a possible
interpretation in terms of smoothing of the accretion pattern,
concentrating on the effect of smoothing on accreting filaments rather
than on gas clumps. The lack of self--similar scaling for the overall
normalisation of the entropy profiles has also been confirmed by Pratt
\& Arnaud (2004) and Piffaretti et al. (2005), based on {\em
  XMM--Newton} observations, and by Mushotzky et al. (2003) from
Chandra data.  Thanks to the possibility to perform spatially resolved
spectroscopy, these authors were able to derive the entropy profiles
from the actual temperature profiles.  Consistent with the results by
Ponman et al. (2003), they concluded that the shape of the entropy
profiles appears to be quite similar for all systems, with an
amplitude that follows the scaling $S\propto T^{2/3}$ more closely
than the $S\propto T$ scaling expected from self--similarity.

The question then arises as to whether the observed entropy structure
of groups and clusters can be used to derive information on the nature
of the feedback that affects the thermodynamical history of cosmic
baryons. Whatever the astrophysical source of this feedback is, it
must act in such a way as to regulate star formation and, at the same
time, to reproduce the observed entropy profiles.

Based on the model of star formation and supernova (SN) feedback by Springel
\& Hernquist (2003a, SH03 hereafter), Borgani et al. (2004) performed a large
cosmological hydrodynamical simulation with the aim of studying the X--ray
properties of galaxy clusters and groups. Although the model provides a
realistic description of the cosmic star formation history (Springel \&
Hernquist 2003b), clusters are found to have approximately self--similar
entropy profiles, which is not consistent with observations. This result
demonstrates that generating the required level of entropy amplification at
cluster outskirts is a challenging task for hydrodynamical simulations even if
they treat star formation and feedback. A number of simple feedback schemes
have been suggested that heat the gas surrounding galaxies in order to better
reproduce the observed entropy scaling (e.g. Kay et al. 2004). While such
experiments provide very useful guidelines, it is clear that a
self--consistent numerical implementation of a physically well motivated
feedback model which can successfully satisfy a large body of observational
constraints is highly desirable.

In this paper, we use hydrodynamical simulations to explore a suite of
different heating schemes, both for radiative and non--radiative runs. We
focus on four objects which cover the mass range from poor clusters to poor
groups.  The aim of our analysis is to verify whether an appreciable entropy
amplification at large radii, as predicted by Voit et al. (2003), can be
obtained with simple preheating schemes and/or with self--consistent models
that include feedback from SN--driven galactic winds.

The main questions that we intend to address with our analysis are the
following: (a) Is the entropy amplification by smoothed accretion confirmed by
hydrodynamical simulations when a simple scenario for pre-heating with an
entropy floor is invoked?  (b) Does this effect persist when simulations
include a realistic treatment for star formation and SN feedback? (c) What are
the observed entropy profiles telling us about the nature of the feedback that
affects the evolution of the ICM and IGM?

The outline of this paper is as follows. In Section 2, we describe the
simulations that we carried out and the different schemes to provide
non--gravitational heating. Section 3 is devoted to the presentation and
discussion of the results, both for the non--radiative and for the radiative
runs. In Section 4, we summarise our results and give our conclusions.

\section{The simulations}
Our simulations have been carried out with {\small GADGET-2}, a new
version of the parallel Tree-SPH code {\small GADGET} (Springel,
Yoshida \& White 2001). It uses an entropy--conserving formulation of
SPH (Springel \& Hernquist 2002), and includes radiative
cooling/heating by a uniform evolving UV background, a sub--resolution
description of star formation from a multiphase interstellar medium,
and the effect of galactic winds powered by SN energy feedback (SH03).

The simulated groups and clusters are the same as in Tornatore et
al. (2003) and Finoguenov et al. (2003), and we refer to these papers
for more details on their characteristics.  Our set of simulated
objects includes a moderately rich cluster with $M_{\rm vir}\simeq
2.6\times 10^{14}h^{-1}{\rm M}_\odot$, and three groups--sized objects
having virial mass in the range $1.6\times 10^{13} \mincir
M_{\rm vir}/(h^{-1}{\rm M}_\odot) \mincir 4.2\times 10^{13}$. The corresponding
halos have been selected from a DM-only simulation of a $\Lambda$CDM
model with $\Omega_m=0.3$, $h=0.7$, $\sigma_8=0.8$ and $f_{\rm bar}=0.13$
for the baryon fraction, within a box of size $70\hm$. As such, they
encompass the mass range characteristic of moderately rich clusters to
groups, where the effect of non-gravitational heating is expected to
be important.

Using a re-simulation technique, the mass and force resolution has
been increased in the Lagrangian regions of the parent box which
correspond by $z=0$ to volumes encompassing several virial radii
around each of the selected halos. The resolution is progressively
degraded in regions farther away, allowing us to save computing time
while still providing an accurate representation of the large-scale
tidal field. In the high resolution region of the cluster simulation,
gas particles have mass $m_{\rm gas}\simeq 2.2\times 10^8 h^{-1} {\rm
M}_\odot$, and we chose $\epsilon_{\rm Pl}=5\,h^{-1}$ kpc for the
Plummer-equivalent softening scale, kept fixed in physical units out
to $z=2$, and then fixed in co-moving units at earlier times. In order
to have a comparable number of particles within the virial regions of
the different simulated structures, we increased the mass resolution
by a factor of eight for the runs of the groups and, correspondingly,
we decreased the force softening by a factor of two. The main
characteristics of the simulations are listed in Table \ref{t:halos}.

We have simulated each object both with radiative and non-radiative
physics. Our non--radiative runs include the case of purely
gravitational heating (GH) and two pre--heating models, which are
implemented by imposing a minimum gas entropy of $S_{\rm fl}=25$ and
100 keV cm$^2$ at redshift $z_h=3$, respectively. Entropy of heated
particles is raised by incresing their internal energy (i.e.,
temperature), while leaving their density unchanged. For the radiative
runs, we have performed several simulations by changing either the
parameters controlling the galactic winds or the level of the
pre--heating entropy floor. In the model of SH03, the velocity of the
galactic winds, $v_w$, scales with the fraction $\eta$ of the SN-II
feedback energy that contributes to the winds, as $v_w\propto
\eta^{1/2}$ (see eq.[28] in SH03). The total energy provided by SN-II
is computed by assuming that they are due to exploding stars with mass
$>8\,{\rm M}_\odot$, each SN releasing $10^{51}$ ergs, and that their
abundance is given by a Salpeter (1955) initial mass function
(IMF). In our simulations, we consider the cases $\eta=0.5$, 1 and 3,
yielding $v_w\simeq 340$, 480 and 840 km s$^{-1}$ (runs W1, W2 and
W3), respectively. While assuming $\eta>1$ is clearly unrealistic in
this picture, it can be phenomenologically interpreted as being due to
an extra energy source, e.g. coming from a top--heavier IMF or from an
AGN component.

In their model for galactic winds, SH03 treated SPH particles that
become part of the wind as temporarily decoupled from hydrodynamical
interactions, in order to allow the wind particles to leave the dense
interstellar medium without disrupting it. This decoupling is
regulated by two parameters. The first parameter, $\rho_{\rm dec}$,
defines the minimum density the wind particles can reach before being
coupled again. If $\rho_{\rm th}$ is the threshold gas density for the
onset of star formation ($\rho_{\rm th}\simeq 2.8\times 10^{-25}$g
cm$^{-3}$ in the SH03 model), then it should be $\rho_{\rm
dec}<\rho_{\rm th}$ for the winds to leave the star-forming region,
and the typical setting by SH03 has been $\rho_{\rm
dec}=0.1-1.0\,\rho_{\rm th}$.  The second parameter, $l_{\rm dec}$,
provides a maximum time via $t_{\rm dec}= l_{\rm dec}/v_e$ a wind
particle may travel freely before becoming hydrodynamically coupled
again. If this time has elapsed, the particle is coupled again, even
if it has not yet reached $\rho_{\rm dec}$. The $l_{\rm dec}$
parameter is introduced only to ensure that the decoupling is stopped
quickly also in cases where the wind particle cannot escape the dense
ISM due to gravity alone.  We provide the values of $\rho_{\rm dec}$
and $l_{\rm dec}$ used for the radiative runs in Table
\ref{t:simul}. The W1, W2 and W3 models all have the same values for
these parameters, but they differ in their wind velocity. In an
attempt to produce more wodely distributed by winds, we have also
run an additional model W4 which has the same wind speed as W3, but
where we allowed the winds to reach much lower densities and larger
distances while being decoupled.

\begin{table}
\centering
\caption{Physical characteristics and numerical parameters of the
simulated halos. Column 2: total mass within the virial radius at
$z=0$ ($10^{13}h^{-1}{\rm M}_\odot$); Column 3: virial radius ($\hm$);
Column 4: mass--weighted temperature within $R_{\rm vir}$ (keV);
Column 5: mass of the gas particles in the simulations
($10^{8}h^{-1}{\rm M}_\odot$); Column 6: Plummer-equivalent
gravitational softening at $z=0$ ($h^{-1}$kpc). The numbers reported
in Columns 2-4 refer to the non--radiative run with gravitational
heating (GH) only, but their values do not significantly change
for the other runs.}
\begin{tabular}{lccccc}
Run & $M_{\rm vir}$ & $R_{\rm vir}$ & $T_{\rm vir}$ & $m_{\rm gas}$ & $\epsilon$\\
\hline 
Cluster & 26.4 & 1.31 & 2.04 & 2.17  & 5.0 \\ 
Group-1 & 4.17 & 0.71 & 0.52 & 0.27  & 2.5 \\ 
Group-2 & 1.75 & 0.53 & 0.39 & 0.27  & 2.5 \\ 
Group-3 & 1.64 & 0.52 & 0.28 & 0.27  & 2.5 \\
\hline
\end{tabular}
\label{t:halos}
\end{table}

\begin{figure*}
\centerline{
\hbox{
\psfig{file=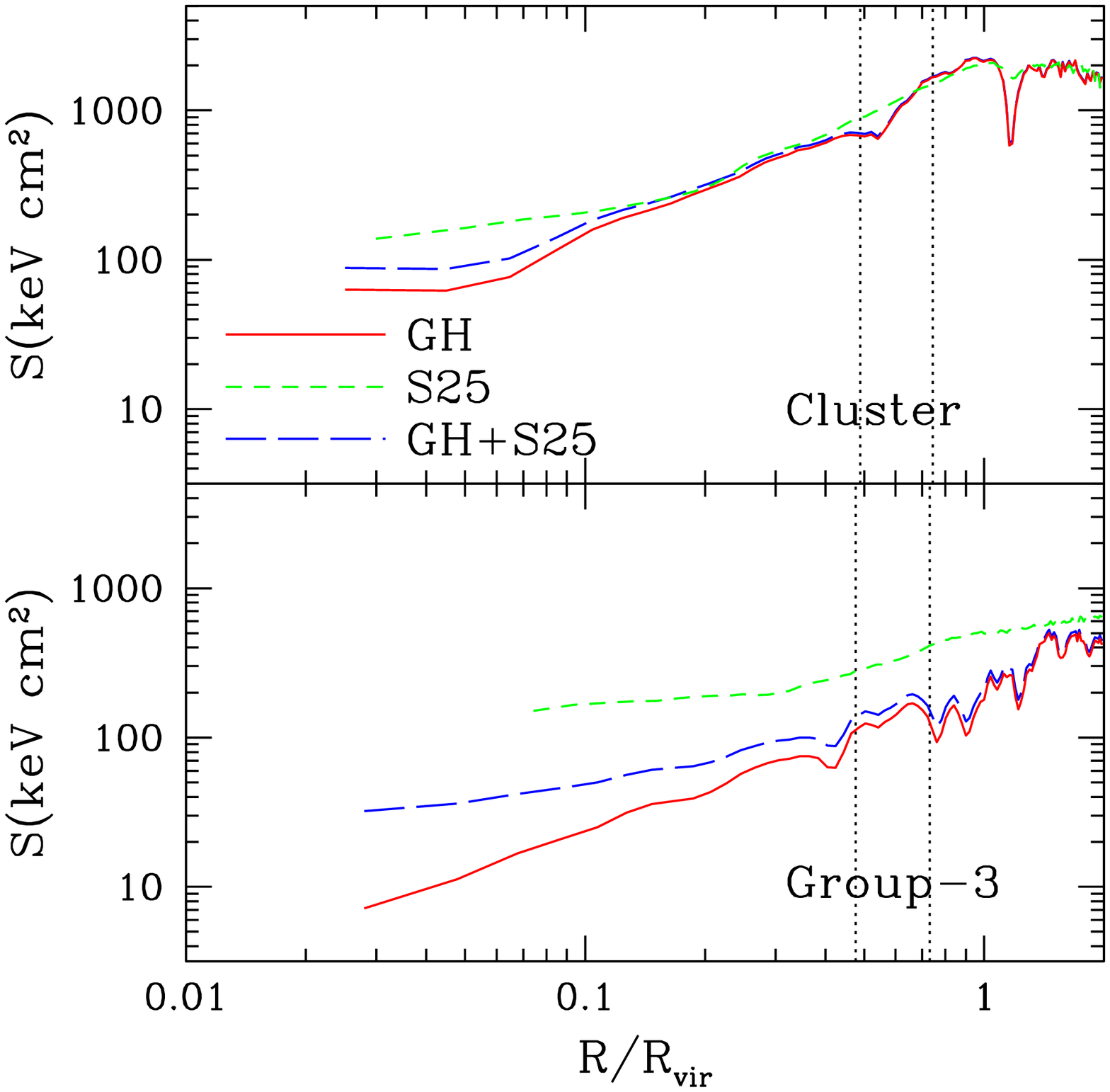,width=8.15cm} 
\psfig{file=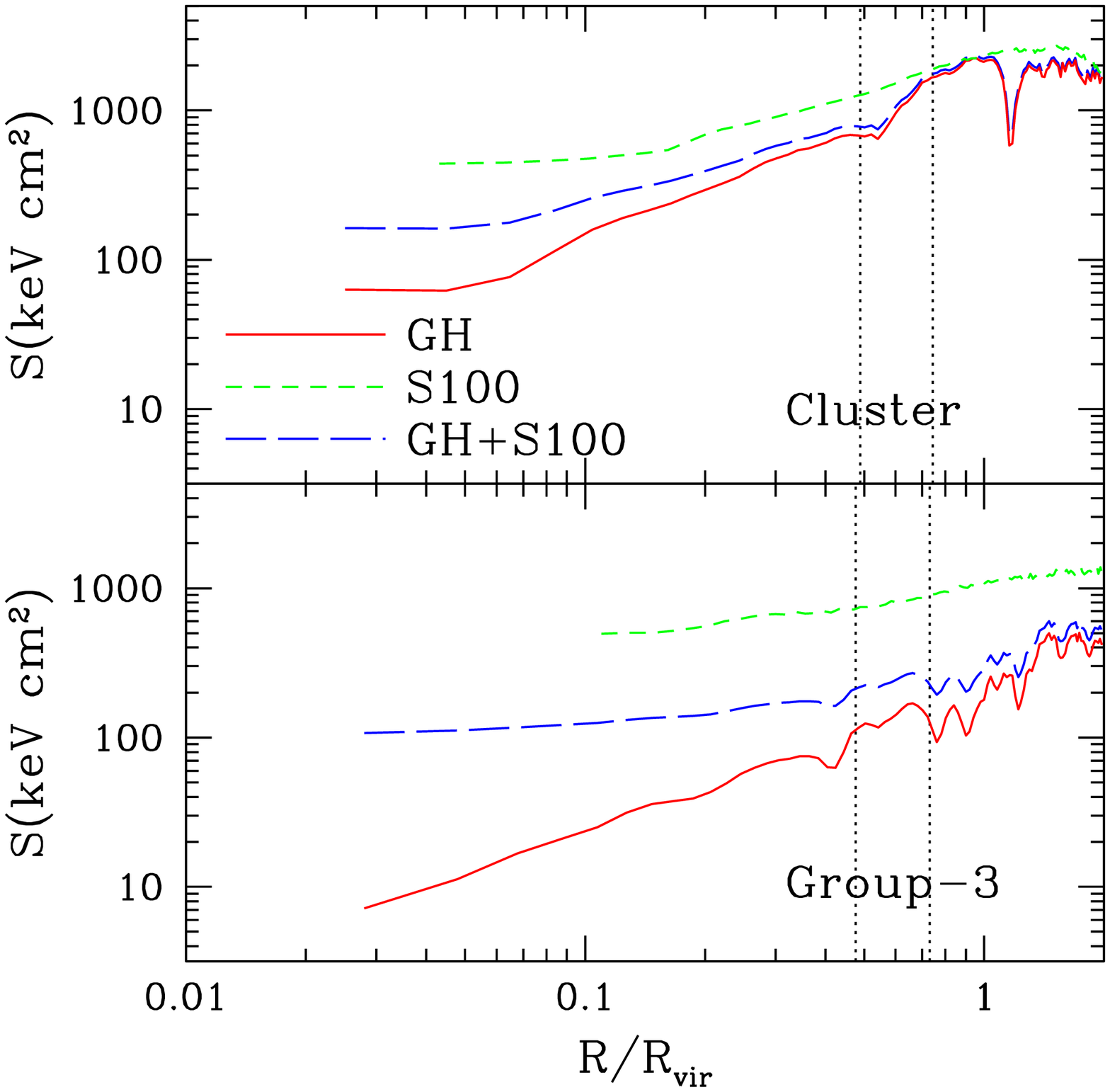,width=8.15cm} 
}
}
\caption{Entropy profiles of the non--radiative simulations. In the
  left and in the right panels, we compare the results for the
  gravitational--heating (GH) runs (solid curves) to those for the
  pre--heated runs (short--dashed curves) with entropy floor $S_{\rm
  fl}=25$ keV cm$^2$ (S25) and $S_{\rm fl}=100$ keV cm$^2$ (S100),
  respectively (see Table 2). Upper and lower panels correspond to the
  simulations of the Cluster and of the Group-3. In each panel, we
  also show with a long--dashed curve the entropy profile obtained by
  adding the entropy--floor value, $S_{\rm fl}$, to the GH case. The
  difference between the short--dashed and the long--dashed curves
  provides the net entropy amplification. The two vertical dotted
  lines mark the positions of $R_{200}$ and $R_{500}$, defined as the
  radii encompassing an average density $200\rho_{\rm cr}$ and
  $500\rho_{\rm cr}$, respectively (here $\rho_{cr}$ is the critical
  cosmic density).}
\label{fi:gh}
\end{figure*}

\begin{table}
\centering
\caption{Description of the different runs. Column 1: run name; Column
2: level of the entropy floor at $z_h=3$ (keV cm$^2$); Column 3: wind
speed (km s$^{-1}$); Column 4: specific heating energy assigned to the
gas particles that fall within $R_{\rm vir}$ at $z=0$ (keV/particle);
Column 5: the limiting density for wind decoupling (in units of the
threshold density, $\rho_{\rm th}$, for star formation (see text);
Column 6: the maximum length that winds travel while being decoupled
(units of $\hk$); Column 7: fraction of baryons in stars within
$R_{\rm vir}$. The values in Columns 4 and 7 refer to the simulations
of the Cluster.}
\begin{tabular}{lcccccc}
Run & $S_{\rm fl}$ & $v_w$ & $E_h$ & $\rho_{\rm dec}$ & $d_{\rm dec}$ & $f_*$ \\
\hline
Non--rad. runs \\
GH  & & & & & &\\
S25 &  25 & & 0.5 & & & \\
S100& 100 & & 2.2 & & & \\
\\
Rad. runs \\
W1 & & 341 & 0.3 & 0.5 & 10 & 0.20 \\
W2 & & 484 & 0.5 & 0.5 & 10 & 0.17 \\
W3 & & 837 & 1.2 & 0.5 & 10 & 0.14 \\
W4 & & 837 & 1.1 & 0.01 & 50 & 0.12 \\
W1+S25 & 25 & 341 & 0.2+0.4$^{\dag}$ & 0.5 & 10 & 0.13 \\
W1+S100 & 100& 341 & 0.2+1.8$^{\dag}$ & 0.5 & 10 & 0.13 \\ 
\hline
\end{tabular}
\begin{enumerate}
\item[$^{\dag}$]{\footnotesize We report separately the energy budget
  associated to galactic winds (see text) and the pre--heating energy
  required at $z=3$ to establish the entropy floor.}
\end{enumerate}
\label{t:simul}
\end{table}

In summary, we have performed 9 runs for each of the four simulated
structures, corresponding to different ways of changing the gas
physics. The characteristics of these runs are given in Table~2. We
also provide in this table the mean value of non--gravitational
specific energy assigned to the gas that falls at the end of the run
within the virial radius of the cluster (comparable amounts are found
for the three groups). For the simulations with pre--heating, we have
determined this energy by tracking back to $z=3$ all the gas particles
within $R_{\rm vir}$ at $z=0$ and summing up the energy contributions
required to increase their entropy to the floor value where needed.
For the runs including SN feedback, we measure the total stellar mass
within $R_{\rm vir}$ at $z=0$ and we compute the total number of SN-II
expected for the assumed Salpeter IMF. The resulting energy is then
multiplied by the efficiency parameter $\eta$ to compute the total
energy associated with wind feedback.

Looking at Table \ref{t:simul}, we note that the heating energy
required to create the entropy floors is comparable to those of the
winds. However, the way in which these two feedback schemes affect the
thermodynamics of the diffuse baryons is intrinsically very
different. The first heating mechanism provides a diffuse energy input
that can in principle reach all the gas particles in the simulations,
and is provided in an impulsive way (i.e. all the energy is released
at the heating redshift $z_h=3$). The second mechanism, instead,
releases energy gradually in time, since it follows the pattern of
star formation, and it is narrowly concentrated around star--forming
regions. As we will discuss in the following, these differences have
an important impact on the smoothing of the accretion pattern and,
therefore, on the entropy generation by accretion shocks.

\section{Results}

\subsection{Non--radiative runs}

In Figure \ref{fi:gh}, we compare the effect of imposing the two
different entropy floors on the entropy profiles of the cluster
simulation and of `Group-3'. Here and in the following we plot
profiles to an innermost radius which contains 100 SPH particles. This
radius has been shown to be the smallest one where numerically
converged results for the X-ray luminosity can be obtained (Borgani et
al. 2002).

As expected, the effect of pre--heating is that of increasing the
level of the ICM entropy by a larger amount for the less massive
system, an effect that increases in strength for a higher entropy
floor. The runs with gravitational heating display a number of wiggles
which mark the positions of merging sub--halos carrying low--entropy
gas. These wiggles are erased once the gas is pre--heated. This is
consistent with the expectation that pre--heating destroys the gas
content of halos whose virial temperature is lower than the specific
pre--heating energy. In the cluster run, the entropy increase due to
pre--heating is marginal at the virial radius, thus implying
that the overall ICM thermal energy content within $R_{\rm vir}$ is mainly
determined by the action of gravity.

\begin{figure*}
\centerline{
\psfig{file=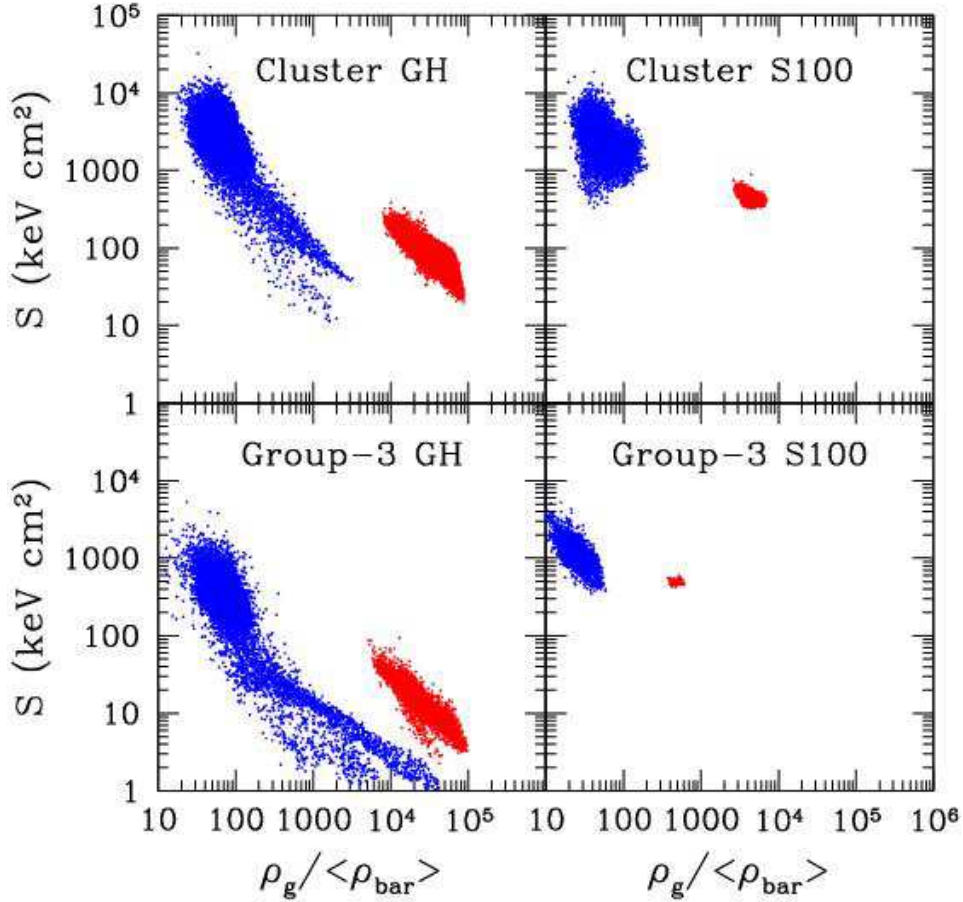,width=13.cm} 
}
\caption{The entropy--density phase diagram for the runs with
  gravitational heating only (GH, left panels) and with pre--heating
  with an entropy floor of 100 keV cm$^2$ (S100, right panel). Upper
  and lower panels refer to the Cluster and to the Group-3,
  respectively. The gas density is given in units of the mean density
  of cosmic baryons. The concentration of (red) points in the lower
  right part of each panel is for the gas particles within
  $0.1\,R_{\rm vir}$, while the concentration of (blue) points
  starting from the upper left part of each panel is for the gas
  particles found between $0.9\, R_{\rm vir}$ and $1.1\, R_{\rm
  vir}$. }
\label{fi:srnr}
\end{figure*}

In Figure \ref{fi:srnr}, we provide further information about the
effect of pre--heating on the entropy structure of the ICM. Here we
show the entropy--density phase diagram of gas particles. The two
separate concentrations of points mark the gas particles falling
within $0.1\,R_{\rm vir}$ and those having cluster-centric distances
in the range $(0.9-1.1)\,R_{\rm vir}$, the former being those having
larger densities. Following these two populations of gas particles
allows us to understand the different effects that non--gravitational
physics has in the innermost and in the outskirt regions of the ICM.
In the outer cluster regions in the GH runs (left panels), the change
of slope in the $S$--$\rho_g$ relation marks the transition from
diffuse (shocked) accretion to clumpy accretion (see also Sect.~3.2
and Figure~\ref{fi:parts} below).
The clumpy accretion is associated with merging subgroups and
filaments, which are characterised by low entropy and high density,
and are erased by pre--heating with an entropy floor of 100 keV cm$^2$
(S100 runs, right panels). This demonstrates that the effects of
pre--heating are: {\em (a)} to cancel any signature of clumpy
accretion in the outskirts; {\em (b)} to suppress the gas density in
the central halo regions, and {\em (c)} to amplify the entropy of both
the diffuse phase in the outer regions and of the gas in the central
regions, by an amount that increases for lower--mass systems.

In their semi--analytic description of entropy generation from diffuse
accretion, Voit et al. (2003) have shown that the post-shock gas
entropy is higher than it would be without preheating, at a given
density, by an amount between $0.84\,S$ and $S$, where $S$ is the
added entropy. Therefore, when comparing the profiles of preheated
simulations to the corresponding GH profiles where the preheating
entropy level is simply added to it, any excess entropy should then be
attributed to the amplification effect due to accretion of smoothed
gas instead of clumps. We expect this effect to become more prominent
for stronger pre--heating, which makes the accretion more diffuse, and
for smaller groups, where the accretion is dominated by comparatively
smaller gas clumps. In Fig.~\ref{fi:gh}, the long--dashed curves show
the profiles obtained by adding the floor value, $S_{\rm fl}$, to the
GH profile. Consistent with the expectation from the model by Voit et
al. (2003), we find clear evidence for the entropy amplification
effect. It is more apparent for the higher entropy floor and for the
smaller halo of the Group-3. For instance, at $R=0.1R_{\rm vir}$, we
obtain for the cluster simulations an amplification of 14 per cent and
84 per cent for the S25 and S100 runs, respectively. These numbers
increase to about 230 per cent and 300 per cent, respectively, for the
`Group-3' runs.

We note that our pre--heating scheme does not increase the entropy of
each gas particle by a fixed extra amount. Instead, it creates an
entropy floor, so that the extra entropy assigned to each particle can
be zero if the particle is already at high entropy, or just the value
required to attain the floor value. On the other hand, the model by
Voit et al. (2003) is a prediction for the entropy amplification when
pre--heating is implemented by adding a constant amount of entropy to
all gas particles. In this case, we would have obtained a stronger
pre--heating and, therefore, an even stronger amplification effect.

The signature of differential entropy amplification extending out to
outer regions of groups and clusters is in line with observational
evidences based both on ROSAT/ASCA (e.g. Ponman et al. 2003) and
XMM--Newton data (e.g. Pratt \& Arnaud 2004; Piffaretti et
al. 2005). However, despite the trend of progressively higher relative
entropy levels in groups, the observational data also indicates that
the shape of entropy profiles is almost independent of mass, with no
evidence for the presence of large isentropic cores.  This is quite
different from the behaviour seen in our non--radiative simulations,
where an amplification of entropy at the cluster outskirts is obtained
at the price of creating much flatter entropy profiles in groups and
larger isentropic cores than in clusters (see Fig.\ref{fi:gh}).

\begin{figure*}
\centerline{
\hbox{
\psfig{file=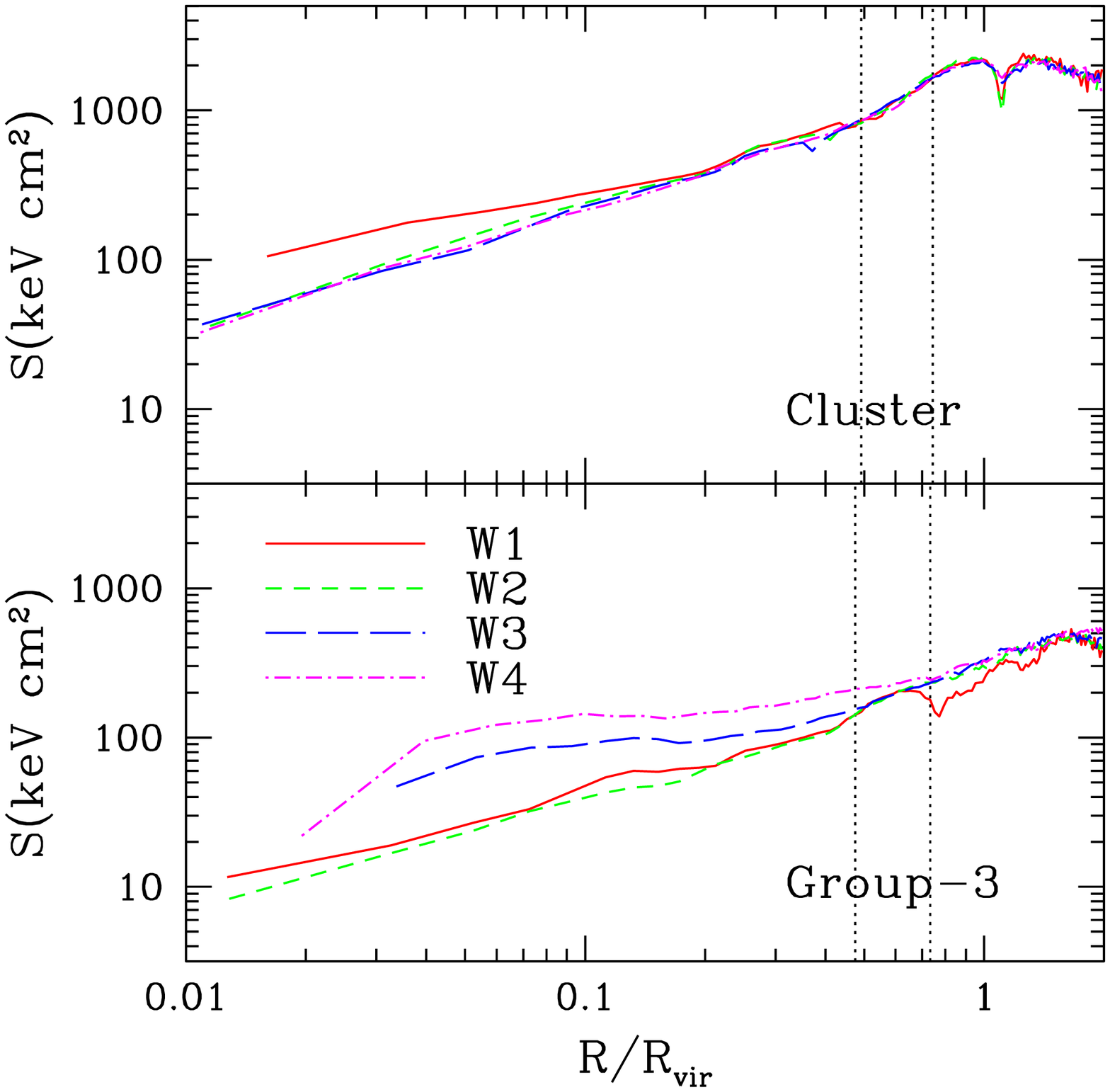,width=8.cm} 
\psfig{file=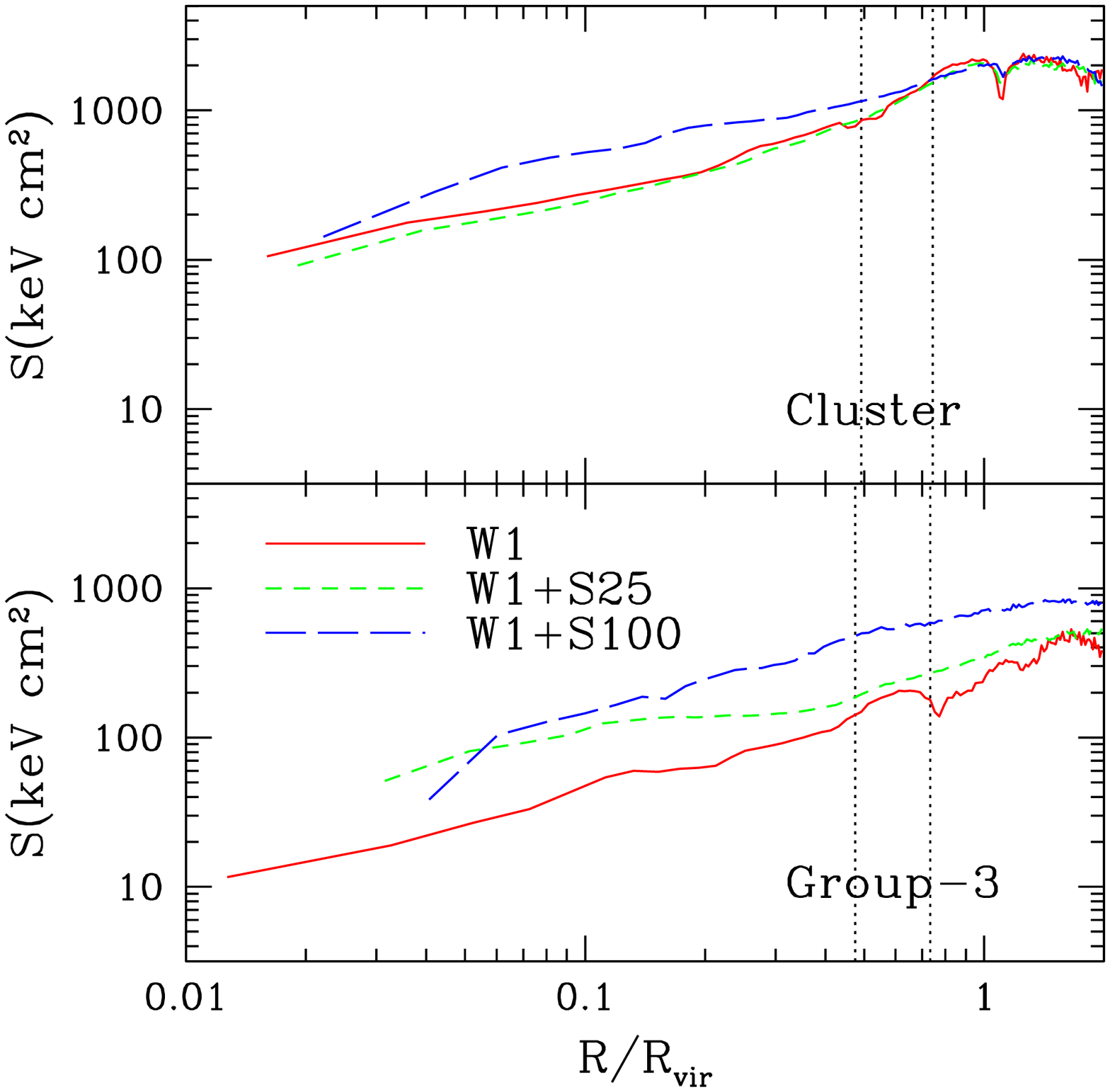,width=8.cm} 
}}
\caption{Entropy profiles of the radiative simulations. The left and
  the right panels show the effect of changing the velocity of the
  galactic winds and of pre--heating with an entropy floor,
  respectively. Results for the Cluster and for the Group-3 are shown
  in the upper and lower panels, respectively. The solid curves always
  indicate the model W1 with $v_w=341\vel$. In the left panels, the
  short--dashed and long--dashed curves are for $v_w=484$ and 720 km
  s$^{-1}$ (W2 and W3 models), respectively, while the dot--dashed
  curves are for the runs with different values for the parameters
  defining the wind decoupling (W4 model, see text). In the right
  panels, short--dashed and long--dashed curves are for the W1 runs
  with entropy floors of 25 and 100 keV cm$^2$ (models W1+S25 and
  W1+S100), respectively. The vertical dotted lines have the same
  meaning as in Fig.\ref{fi:gh}.}
\label{fi:csf} 
\end{figure*}

\subsection{Simulations with radiative cooling}

The results for the entropy profiles change substantially for the
radiative runs. In Figure \ref{fi:csf}, we compare the profiles for
the simulations including cooling and star formation, obtained by
either changing the parameters of the winds (left panels) or by adding
a pre--heating entropy floor to the feedback of the W1 model. As for
the ``Cluster'', increasing the wind speed leaves the entropy profiles
completely unaffected in the halo outskirts.  As for the central
regions, increasing the wind speed from the W1 to the W2 model
slightly decreases the entropy level. In fact, the stronger feedback
creates a population of lower entropy particles which are kept in the
hot phase by the larger amount of energy feeback, instead of cooling
down. Further increasing the wind speed (W2 and W3 runs) or changing
the parameters defining the wind decoupling do not further change the
entropy profile for the Cluster. 

Although the effect of stronger winds
is more apparent for the `Group-3' run, it is nevertheless still
marginal for $R\magcir 0.5\,R_{\rm vir}$.  However, increasing the
wind speed has a significant effect on the entropy profile in the
inner regions of `Group-3': the stronger feedback creates a nearly
isentropic regime for $0.07\,\mincir R/R_{\rm vir}\mincir 0.5$, while
the profile steepens again in the innermost regions, where cooling
starts to dominate. Quite intriguingly, a similar behaviour has been
recently found by Mahdavi et al. (2005), based on an analysis of
galaxy groups observed with XMM--Newton. They find entropy profiles
with a broken power--law shape $S\propto R^\alpha$, with an inner and
outer slope of $\simeq 0.9$ and $\simeq 0.4$, respectively, and a
transition that takes place at about $0.1R_{500}$.  We also note that
the presence of a strong wind reduces the resulting fraction of stars
from about 20 per cent to about 13 per cent (see Table 2). On one hand
this demonstrates that strong winds do increase entropy and,
therefore, the cooling time of the gas surrounding star--forming
regions (e.g., Springel \& Hernquist 2003b; Borgani et al. 2004, in
preparation). On the other hand, the results shown in the left panels
of Fig.~\ref{fi:csf} also indicate that this effect is not causing a
transition from clumpy to diffuse accretion, as would be required to
generate an entropy amplification effect.

Allowing the wind particles to be decoupled until they reach lower
densities, and hence allowing them to travel to a larger distance from
the star--forming regions before interacting with the diffuse gas
(model W4), only produces a modest change. Since the mechanical wind
energy is thermalized at lower gas densities in this model, radiative
losses are reduced as a consequence of longer cooling times.  However,
the resulting additional suppression of star fraction (see Table
\ref{t:simul}) and increase of the entropy in the `Group-3' simulation
are too small to change our conclusions. 

\begin{figure*}
\centerline{
\psfig{file=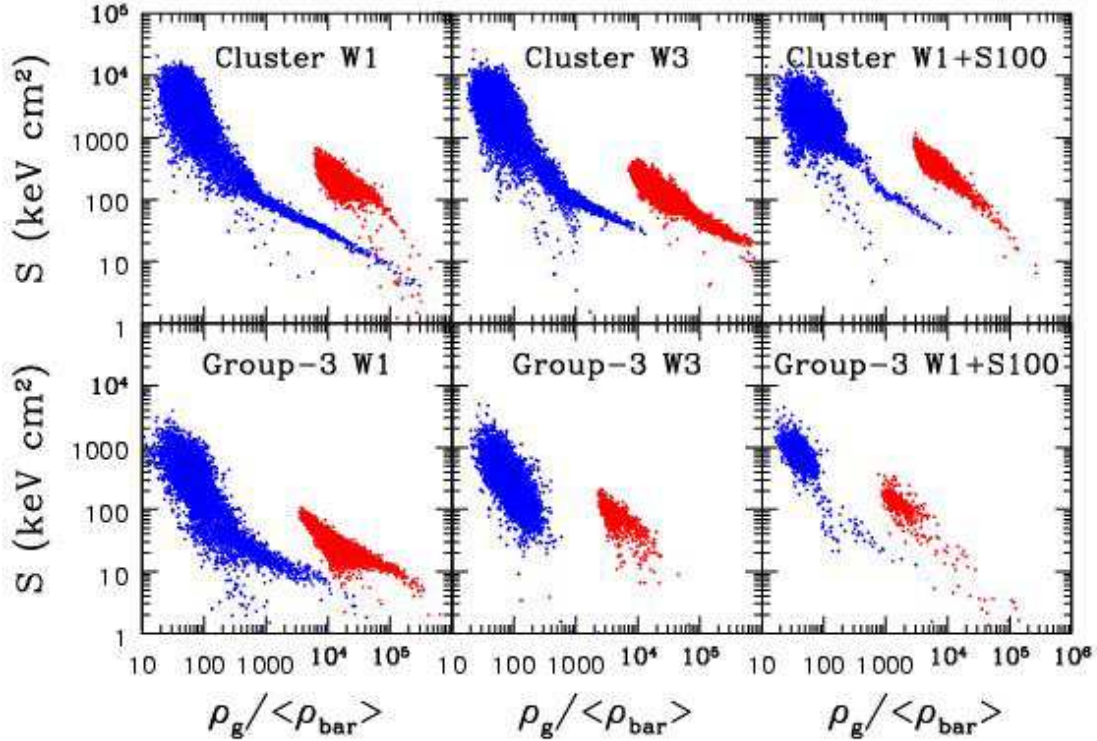,width=15.cm} 
}
\vspace{-4.truecm}
\caption{The same as in Figure \ref{fi:srnr}, but for the radiative
  runs. Left, central and right panels are for the W1, W3 and W1+S100
  runs, respectively.}
\label{fi:srrad}
\end{figure*}

\begin{figure*}
\centerline{
\vbox{
\hbox{
\psfig{file=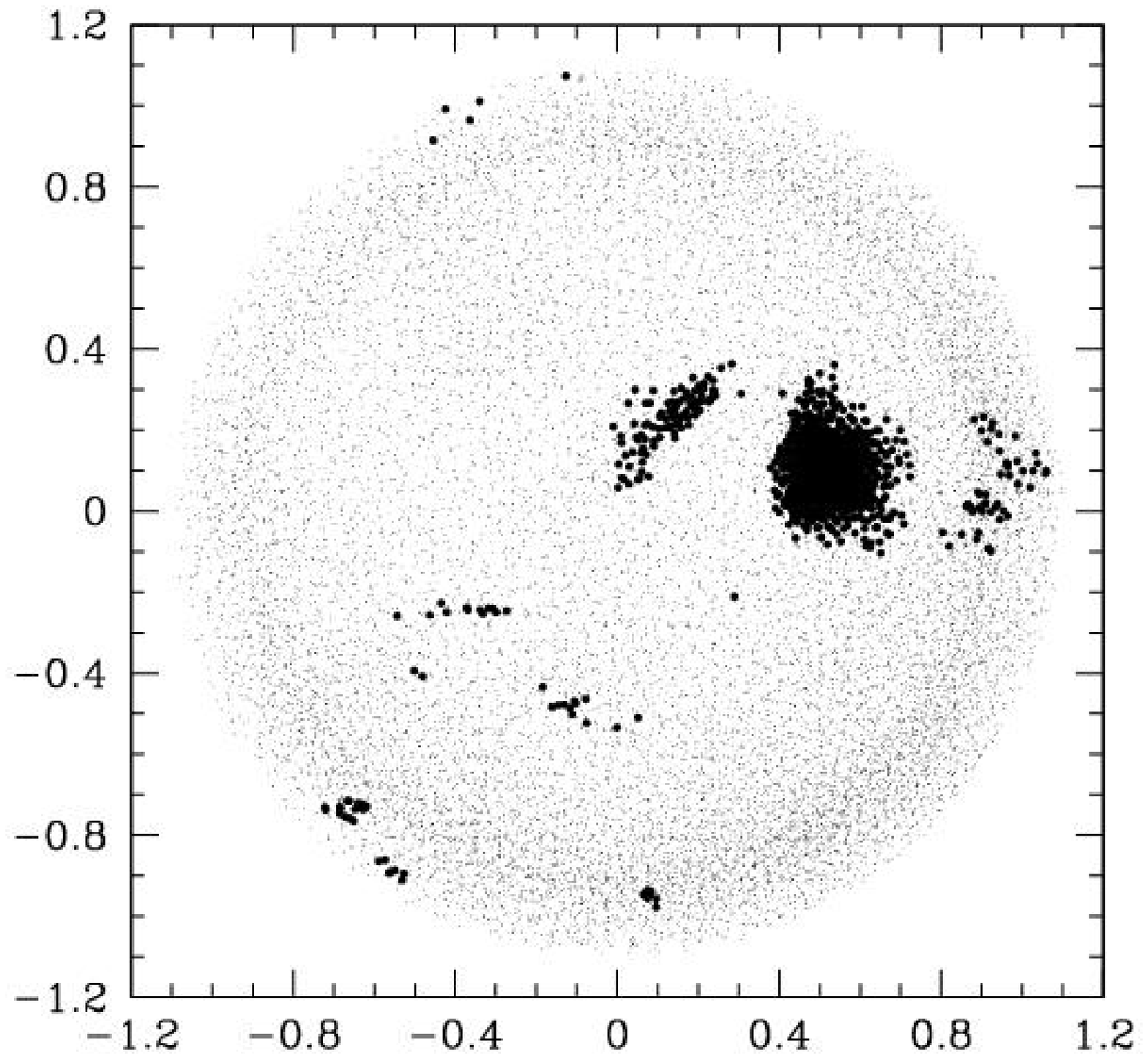,width=6.8cm} 
\hspace{-1truecm}
\psfig{file=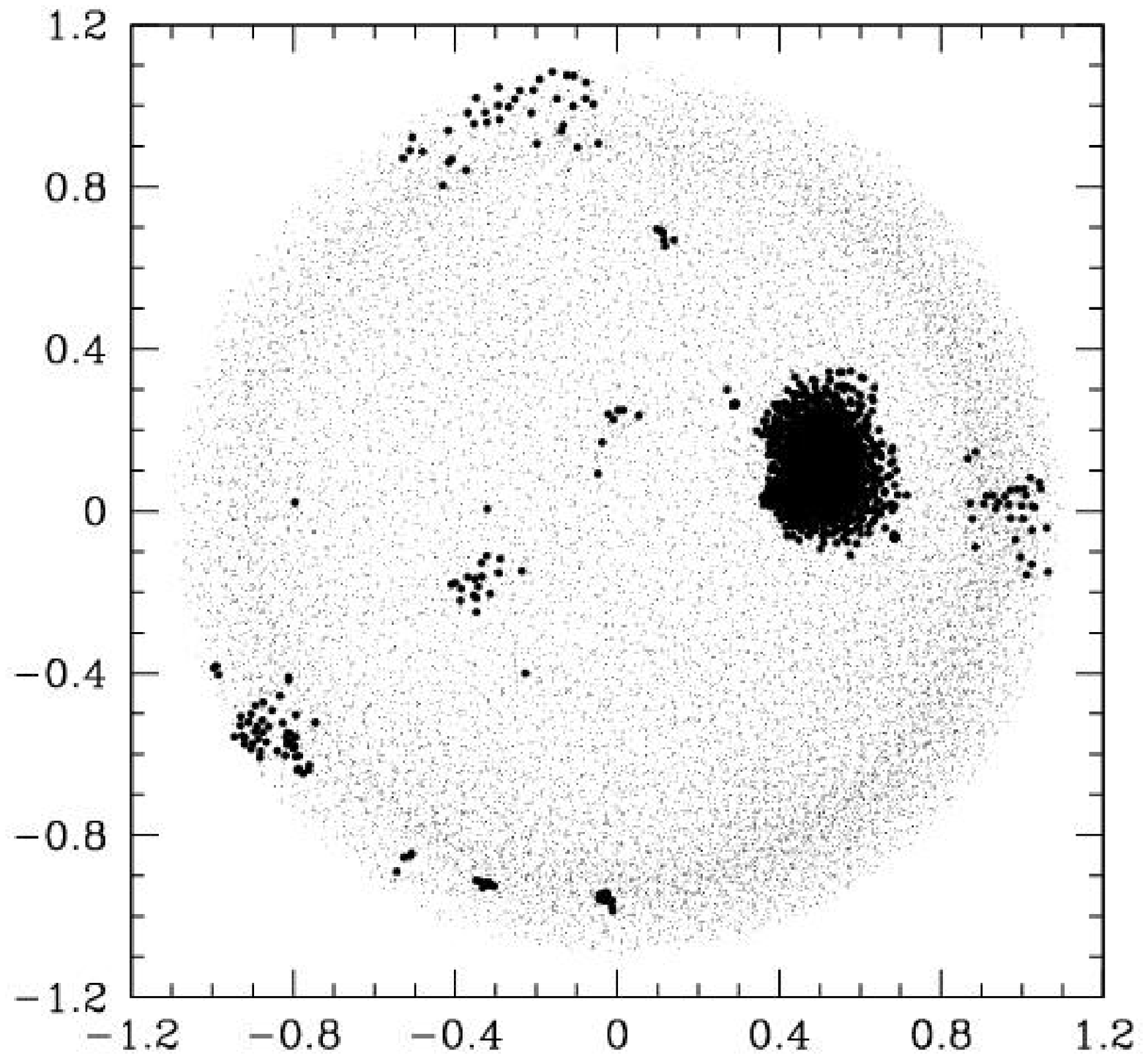,width=6.8cm} 
\hspace{-1truecm}
\psfig{file=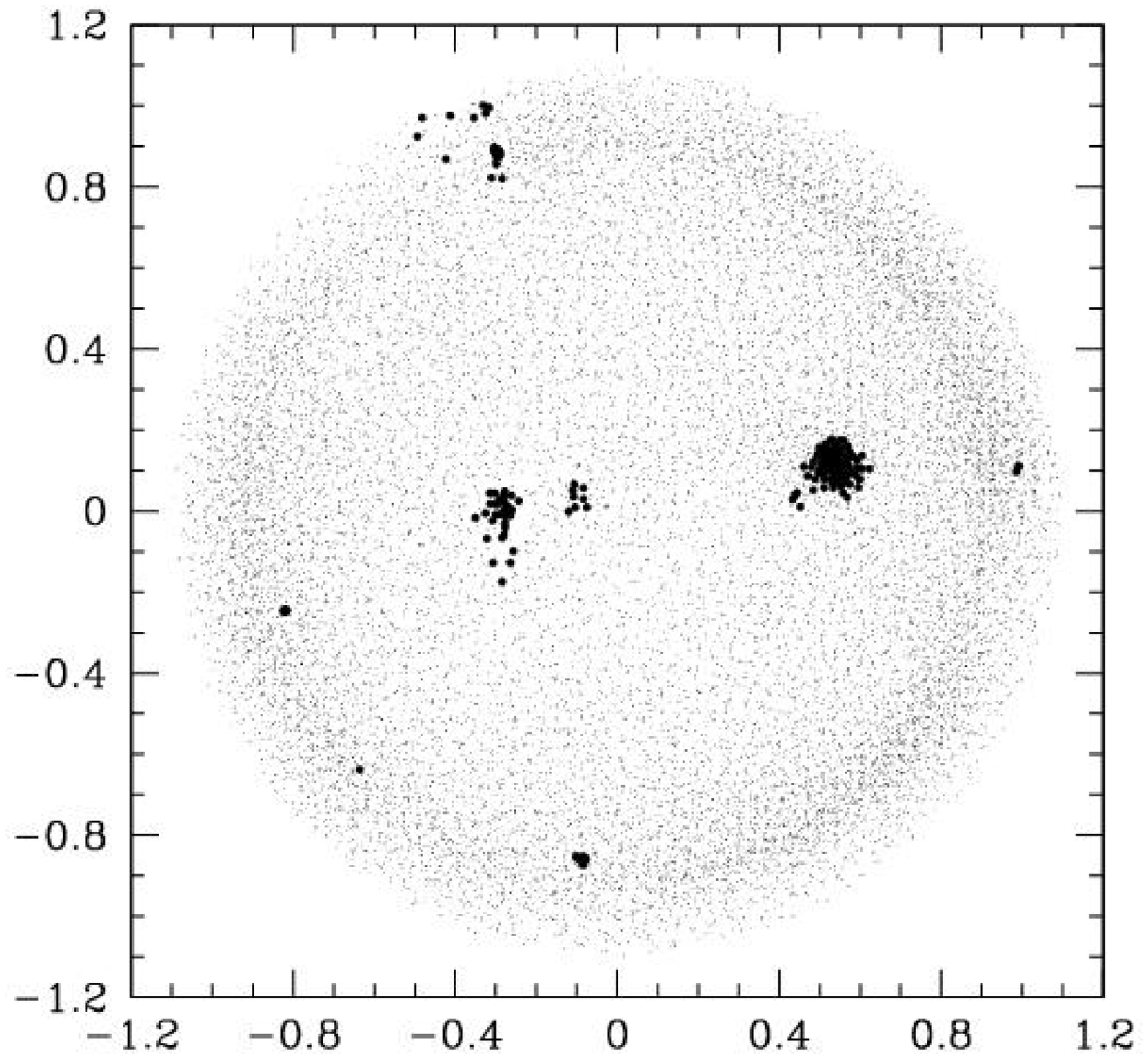,width=6.8cm} 
}
\vspace{-1truecm}
\hbox{
\psfig{file=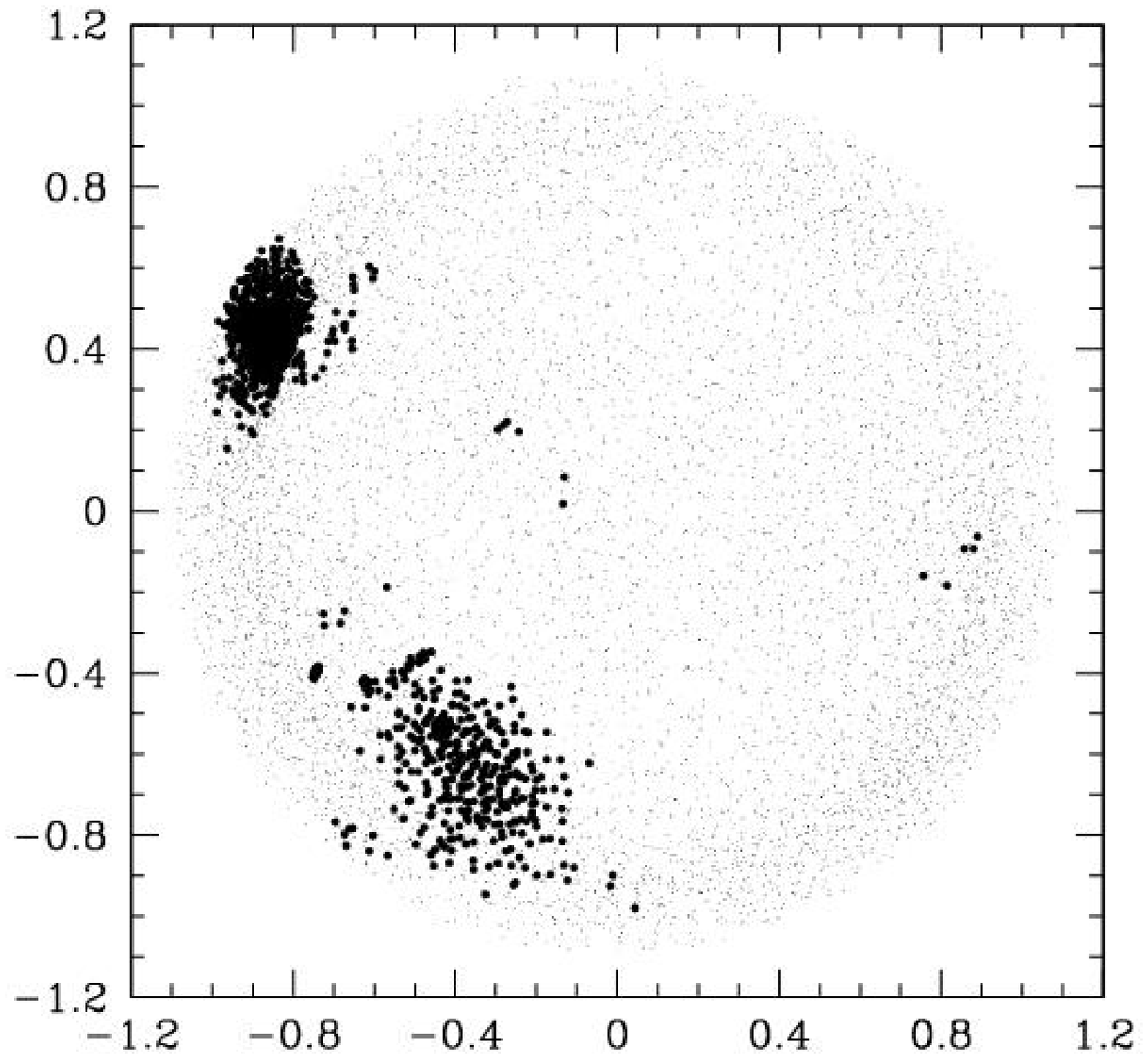,width=6.8cm} 
\hspace{-1truecm}
\psfig{file=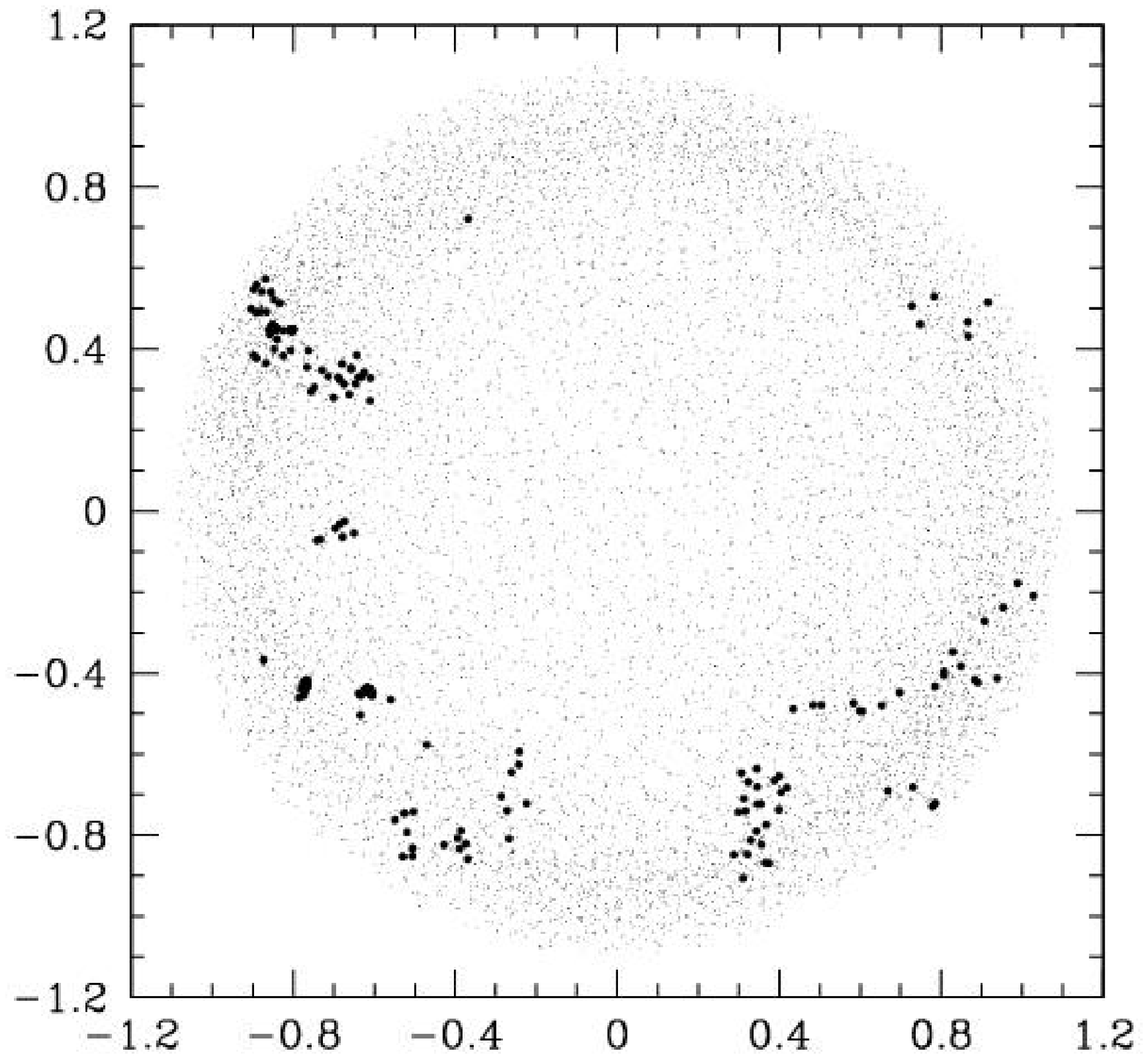,width=6.8cm} 
\hspace{-1truecm}
\psfig{file=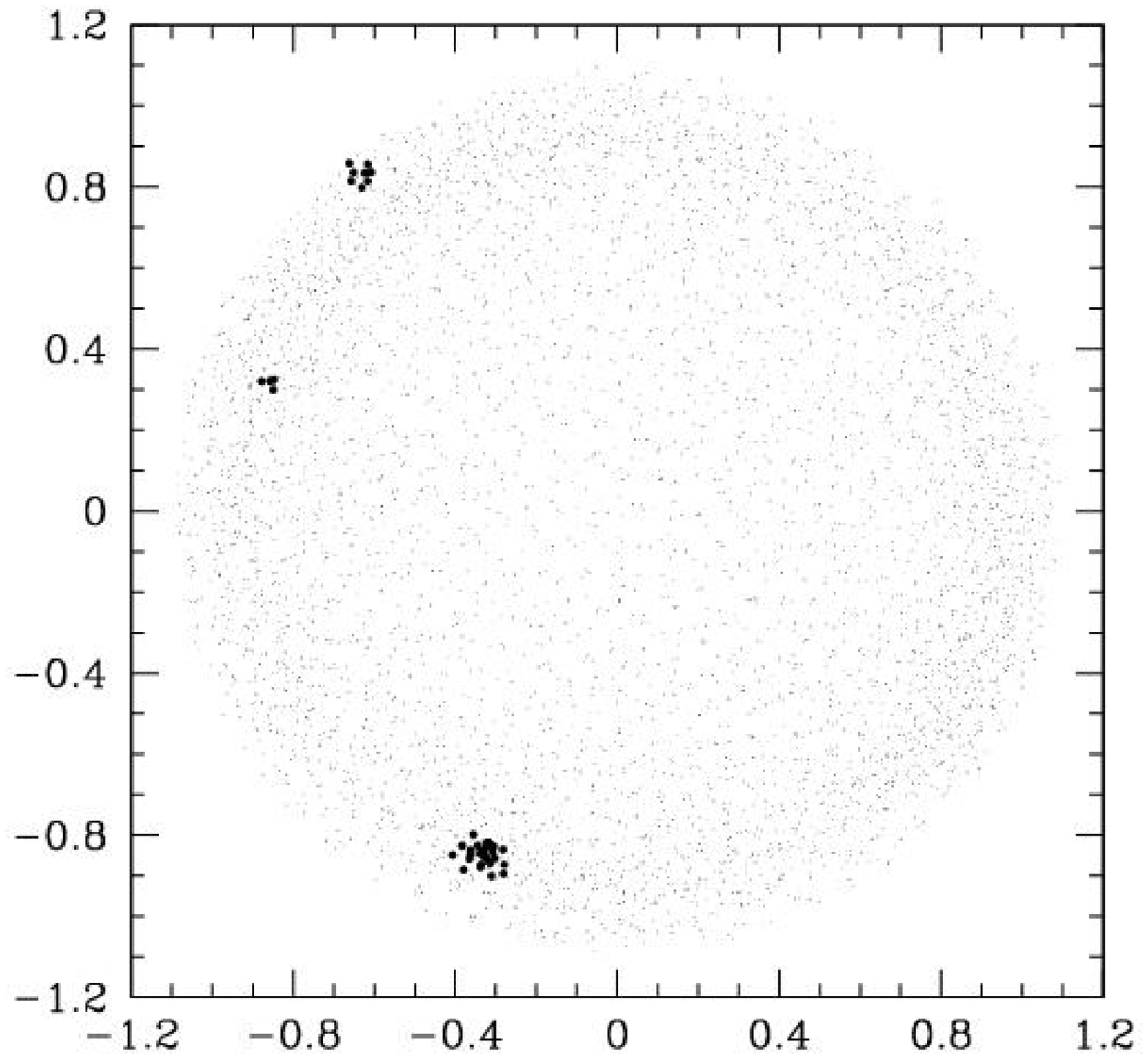,width=6.8cm} 
}
}}
\caption{The distribution of gas particles in the outskirts
  ($0.9<R/R_{\rm vir}<1.1$) of the Cluster (upper panel) and Group-3
  (lower panels). Left, central and right panels are for the W1, W3
  and W1+S100 models, respectively. Light and heavy points are for the
  gas particles whose entropy is above and below the value of 400 keV
  cm$^2$ for the Cluster and of 60 keV cm$^2$ for the Group-3. This
  plot demonstrates that such entropy values approximately mark the
  transition between clumpy and diffuse gas for the two simulated
  structures.}
\label{fi:parts} 
\end{figure*}

\begin{figure*}
\centerline{ \vbox{ \hbox{ \psfig{file=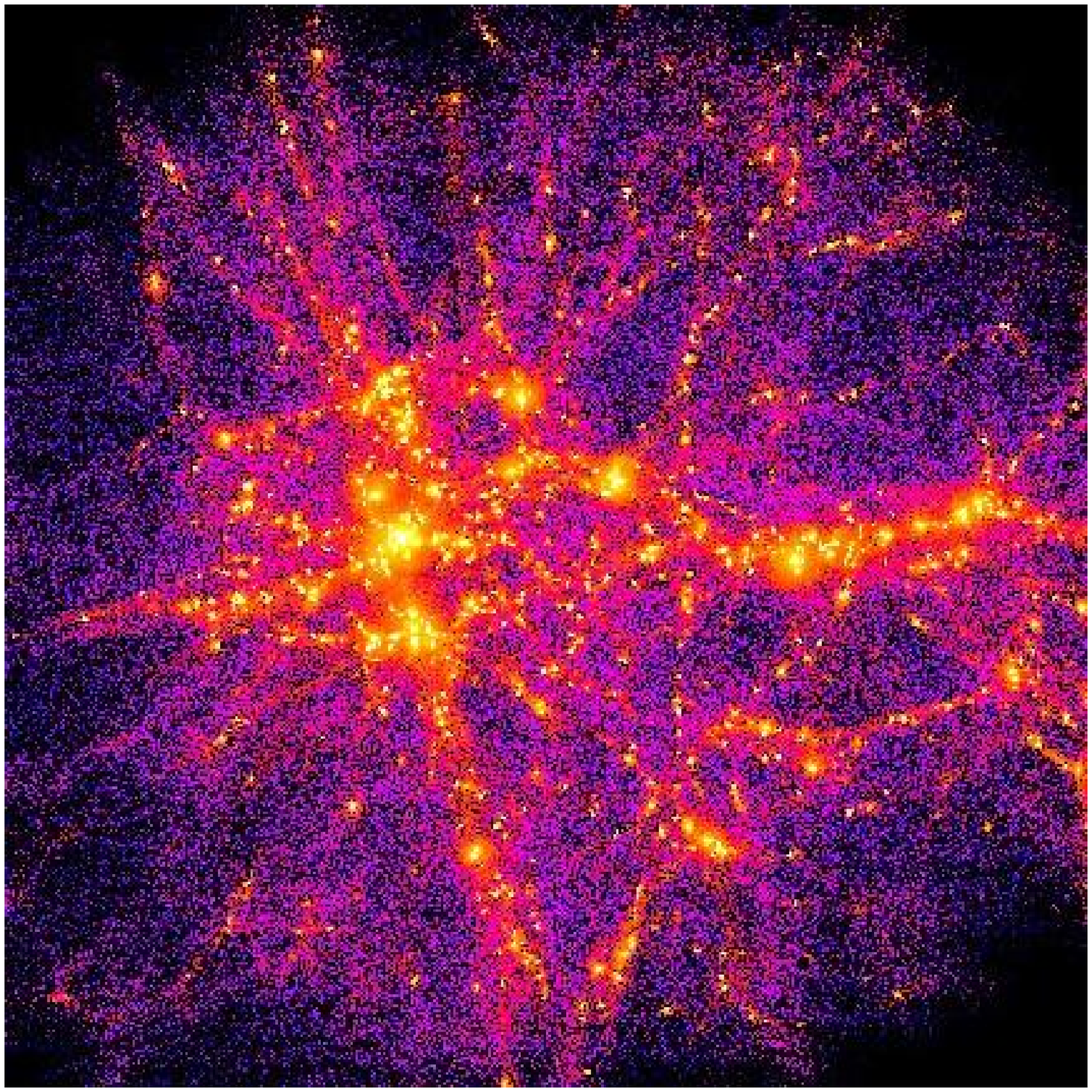,width=7.5cm}
\psfig{file=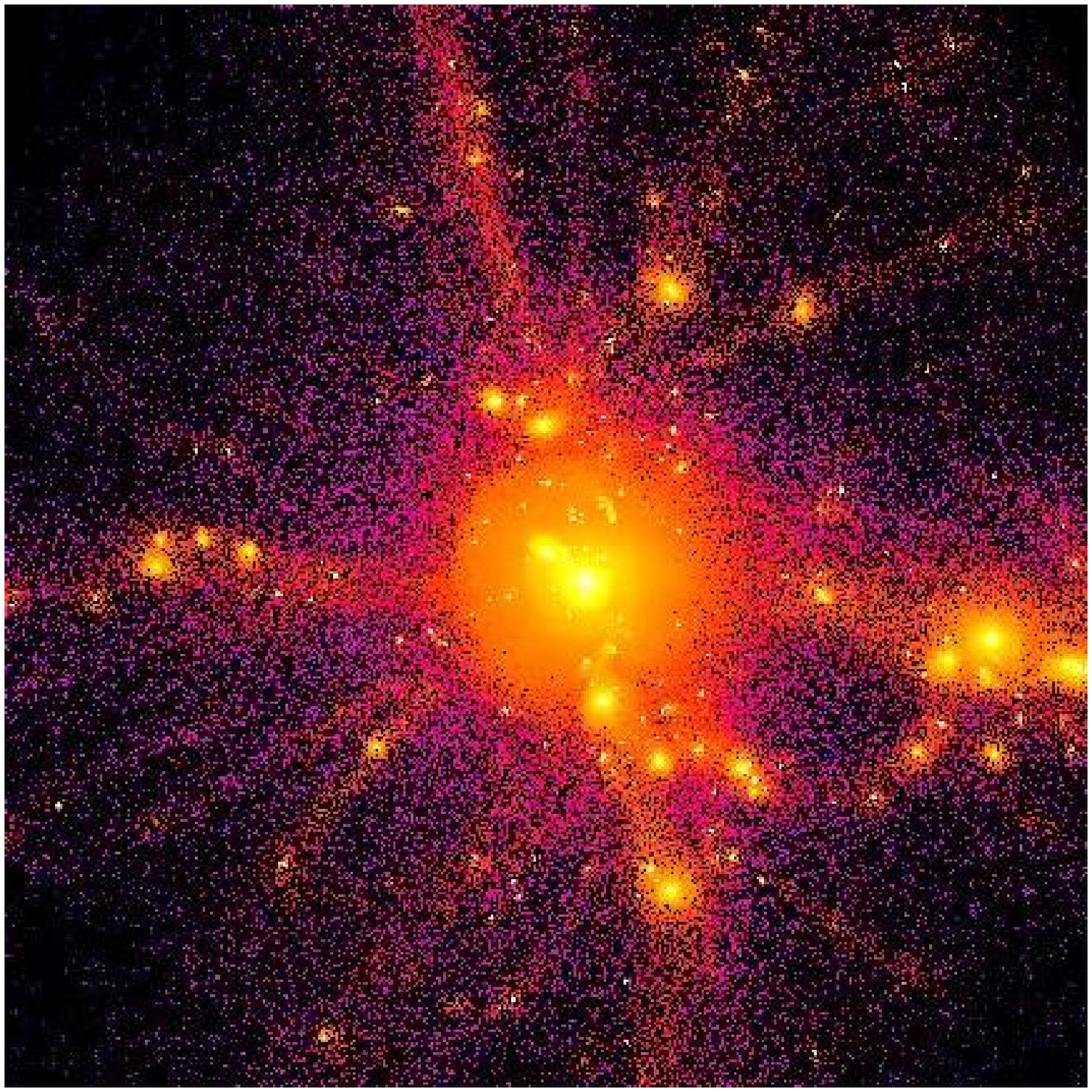,width=7.5cm} } \hbox{
\psfig{file=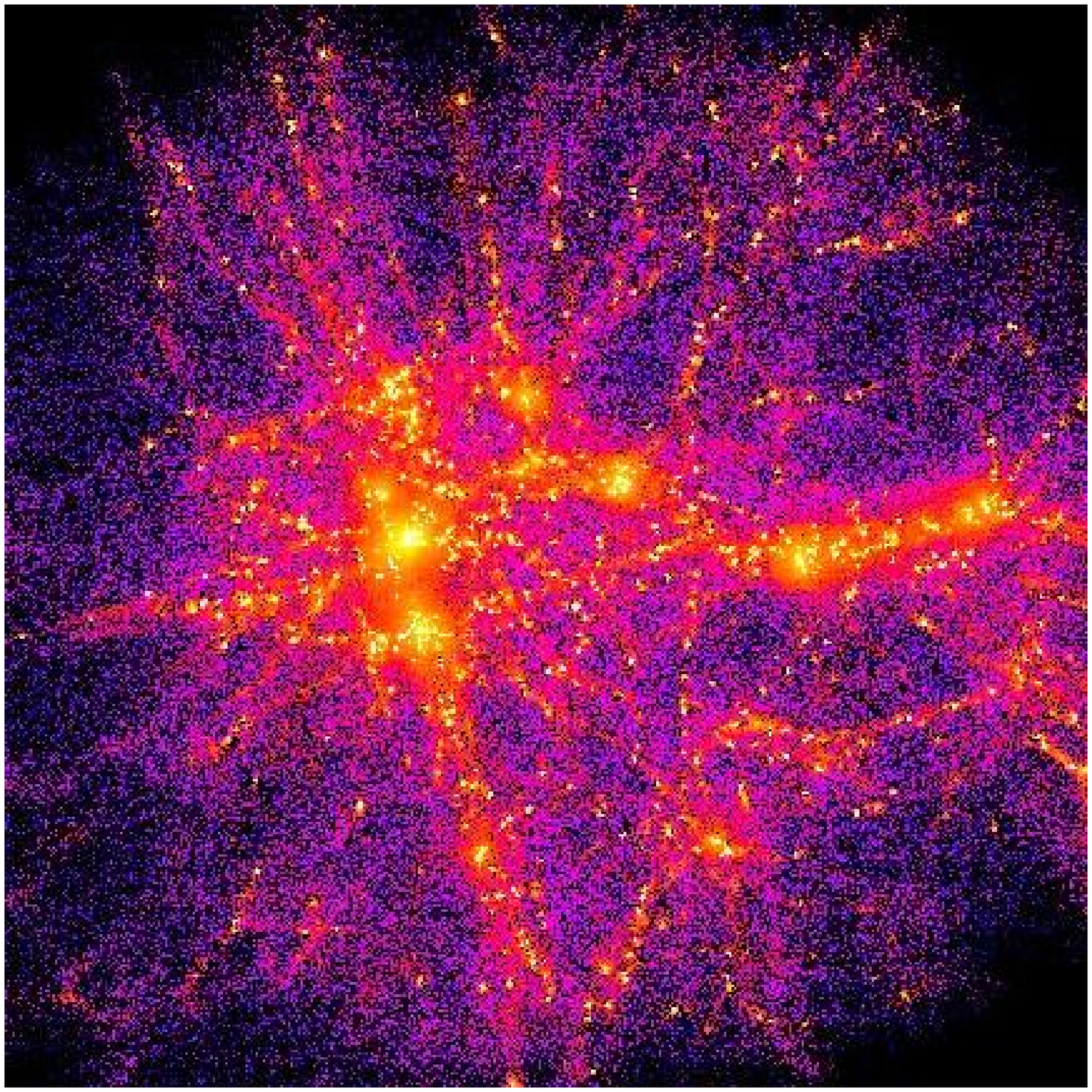,width=7.5cm}
\psfig{file=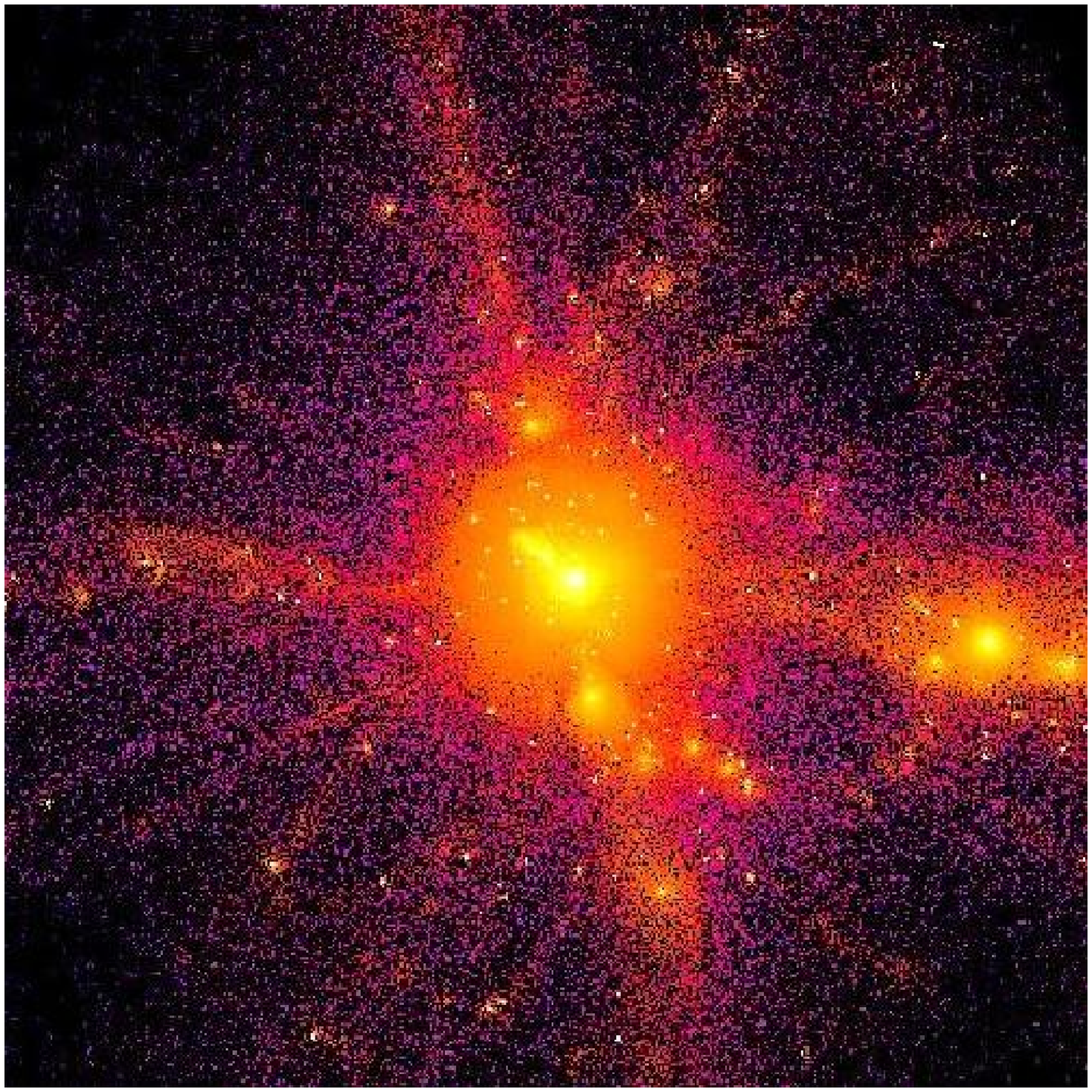,width=7.5cm} } \hbox{
\psfig{file=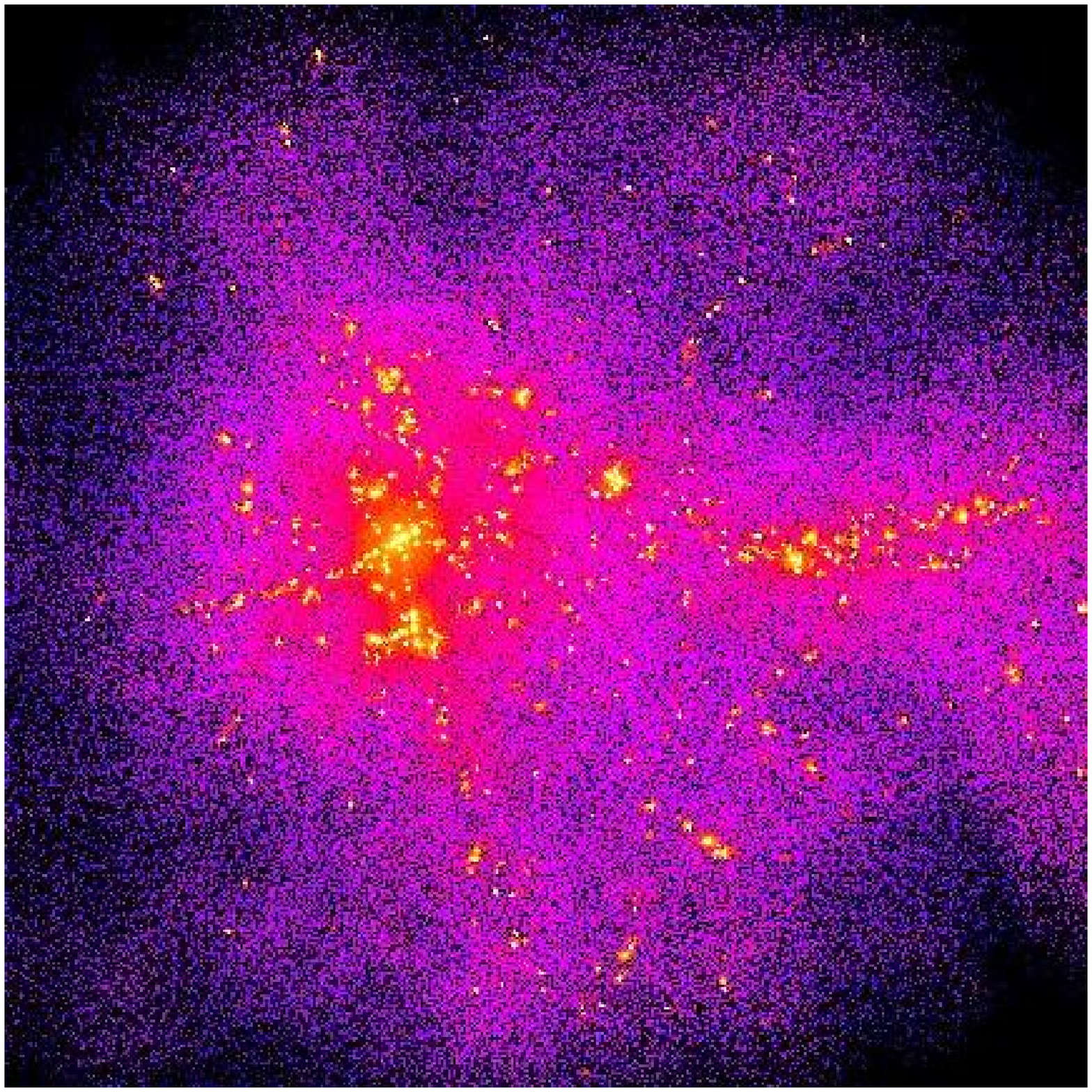,width=7.5cm}
\psfig{file=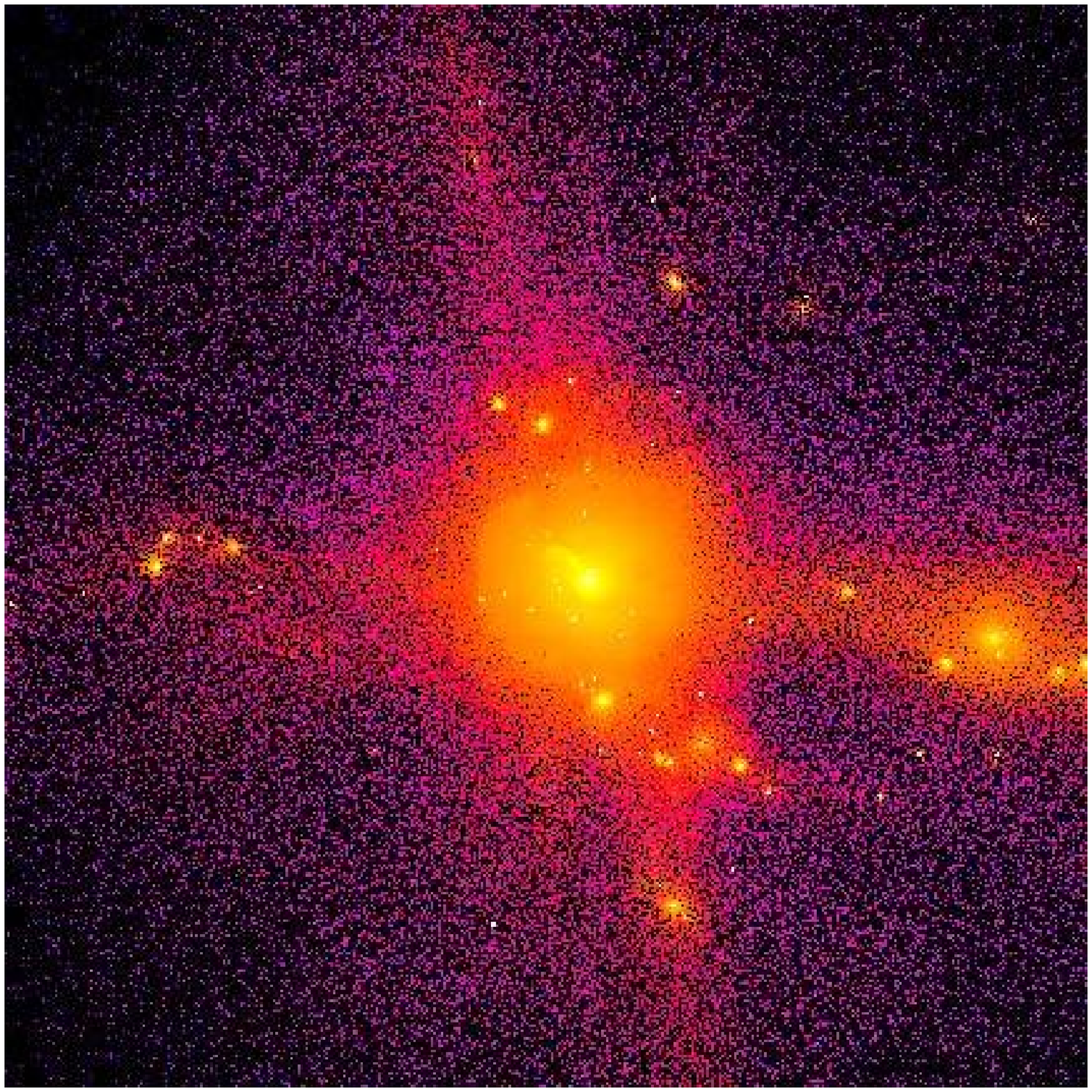,width=7.5cm} } }}
\caption{Maps of the gas density for the radiative runs of the
Cluster. Upper, central and lower panels are for the W1, W3 and
W1+S100 runs, respectively. Left and right panels are for the outputs
at $z=2$ and $z=0$, respectively. At $z=0$ the size of the box is 
$11.7\hm$, while at $z=2$ it corresponds to $17.5\hm$ comoving. The
small white knots mark the ``galaxies'', i.e.~the places where high
density gas is undergoing cooling and star formation.}
\label{fi:maps} 
\end{figure*}

These results on the effect of winds are also supported by the
phase--diagrams shown in Figure~\ref{fi:srrad}, by the relative
distribution of low--entropy and high--entropy particles in the halo
outskirts, shown in Figure~\ref{fi:parts}, and by the maps of gas
density of the `Cluster' runs, shown in
Figure~\ref{fi:maps}. The latter demonstrates that both in the
proto--cluster region at $z=2$ and in the cluster at $z=0$, the
stronger winds wash out only the smallest halos and make the larger
ones slightly puffier, while preserving the general structure of the
cosmic web surrounding the Lagrangian cluster region.

By comparing the results of the W1 and W3 models in
Figs.~\ref{fi:srrad} and \ref{fi:parts} (left and central panels,
respectively), we see that a higher wind speed has also a negligible
effect on the Cluster, despite being able to erase accreting clumps in
the outskirts and to push the gas in the central regions of the
`Group-3' simulation to higher entropy. In all cases, we find that the
stronger winds of the W3 and W4 models are quite ineffective in
increasing the entropy of the diffusely accreting gas.

A larger effect on the ICM entropy structure is induced by adding a
pre--heating entropy floor to the W1 feedback (right panels of
Fig.~\ref{fi:csf}). In this case, a sizable amplification is obtained
for both the `Group-3' and the `Cluster' simulations, at least when
the higher entropy floor, $S_{\rm fl}=100$ keV cm$^2$, is used. As
shown in Fig.~\ref{fi:maps}, this entropy amplification is associated
with a much smoother gas density distribution, both at $z=2$ and at
$z=0$. In this case, the filamentary structure of the gas distribution
is completely erased, while only the largest halos are able to retain
part of their gas content. This visually demonstrates that a
transition from clumpy to smooth accretion during the process of
cluster formation requires a more broadly distributed form of
feedback.

The effect on the $S$--$\rho_g$ phase diagram is shown in the right
panels of Fig.~\ref{fi:srrad}: Adding the entropy floor has a stronger
impact in smoothing the pattern of gas accretion than increasing the
wind speed. Also, Fig.~\ref{fi:parts} shows that almost no signature
of clumpy accretion is left in the `Group-3' run. Comparing these
results to those of the non--radiative runs shows that radiative
cooling significantly reduces the entropy amplification. The main
reason for this is that cooling increases gas density within accreting
clumps and filaments, thus partially inhibiting the transition from
clumpy to diffuse accretion. This effect can be appreciated by
comparing the left panels of Fig.~\ref{fi:srrad} with the left panels
of Fig.~\ref{fi:srnr}. While the presence of cooling does not
significantly change the entropy level of the accreting diffuse gas,
it nevertheless allows residual low--entropy clumps in the outskirts
to survive the pre--heating, thus lowering the overall amplitude of
the entropy profiles.

An interesting characterisation of the effect of either increasing the
wind speed or adding an entropy floor is obtained by comparing the
star fraction with the corresponding specific extra energy involved by
non--gravitational heating (see Table 2). Increasing the wind speed in
the `Cluster' runs has only a marginal effect on the entropy profiles,
but it is quite effective in regulating cooling: the W3 model produces
$\sim 30$ per cent fewer stars than the W1 model, bringing it into
better agreement with observational data (e.g., Lin, Mohr \& Stanford
2003) which generally favour low stellar densities. Quite
interestingly, strong winds are at least as efficient as the S100
model in preventing overcooling, while requiring a lower amount of
extra energy per gas particle. This highlights the different ways in
which winds and an entropy floor affect the thermodynamics of the diffuse
baryons. The former acts locally to increase the cooling time of the
gas surrounding the star forming regions but it is inefficient in
smoothing out the gas content of merging sub--halos. The latter
provides a diffuse impulsive heating and provides for a much more
efficient transition from smooth to diffuse accretion.

\begin{figure*}
\centerline{
\hbox{
\psfig{file=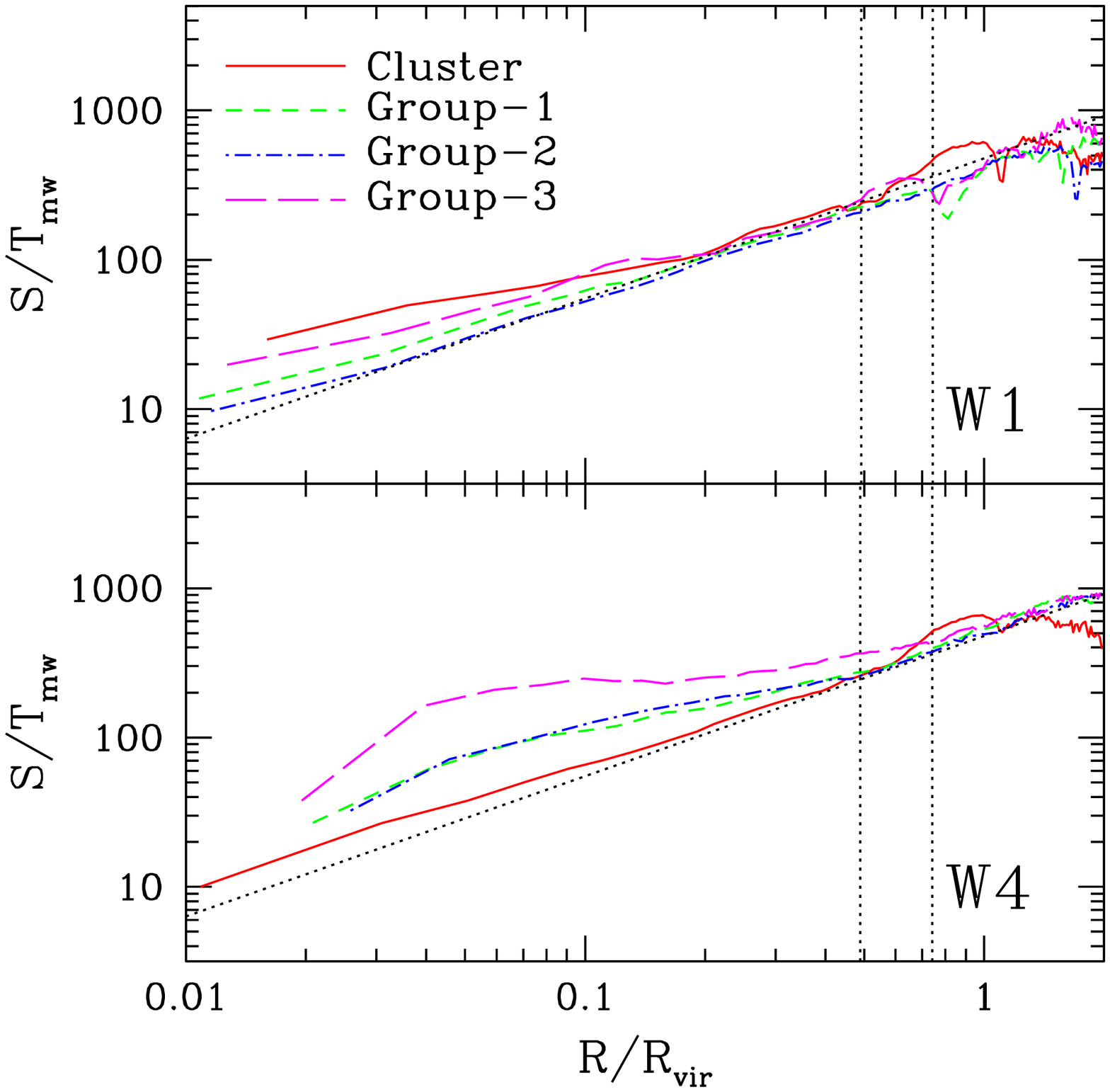,width=8cm} 
\psfig{file=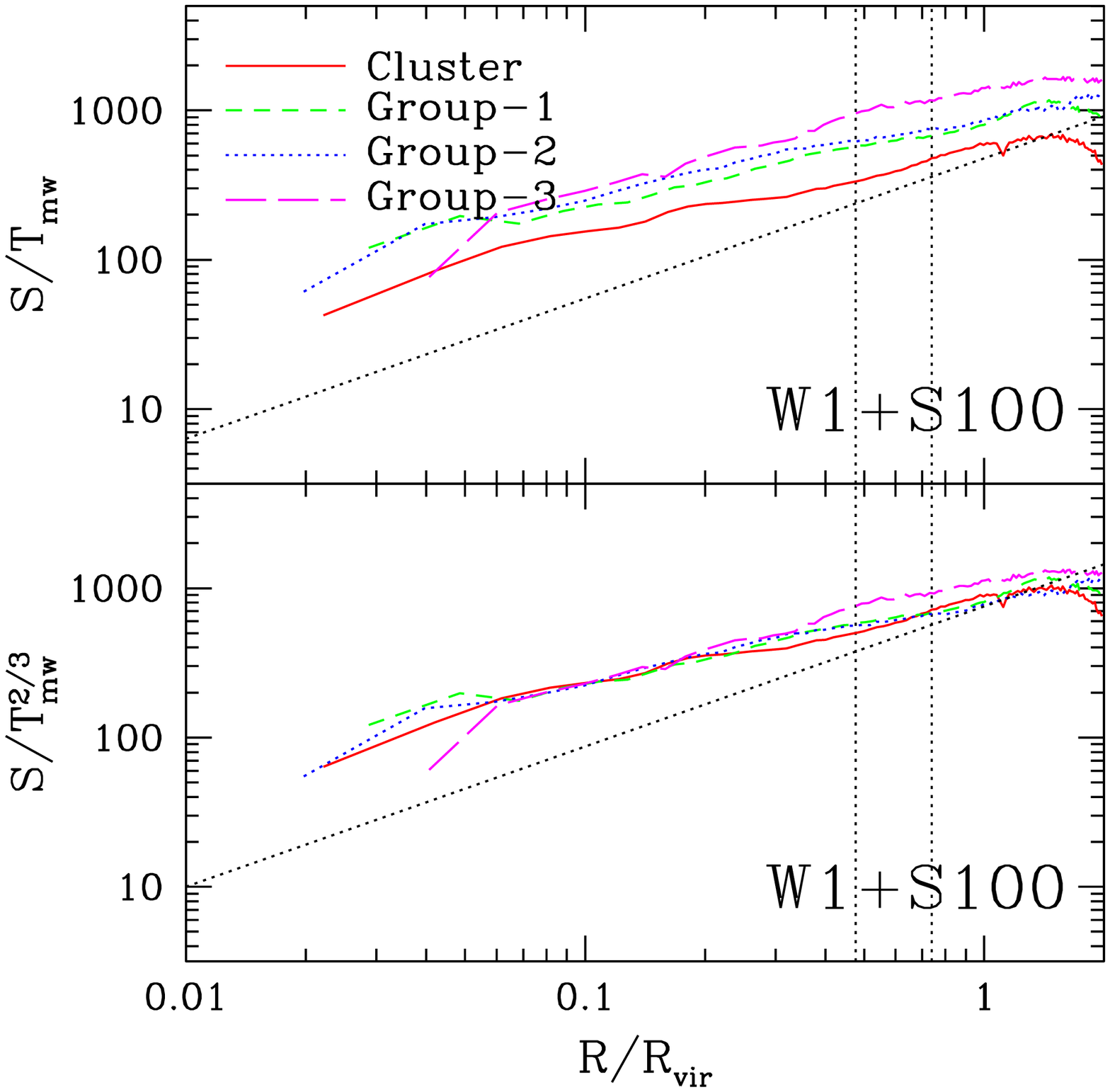,width=8cm} 
}}
\caption{The profile of reduced entropy for the radiative runs. The
  left panels and the upper right panel show the results for the runs
  with wind velocity $v_W=341\vel$ (W1 runs) and $837\vel$ (W4 runs),
  and for the W1 runs that also include an entropy floor of 100 keV
  cm$^2$ (W1+S100), respectively. The profiles are scaled with the
  mass--weighted temperature, $T_{\rm mw}$. The lower right panel
  shows the results of the W1+S100 runs, but with entropy profiles
  normalised according to the observed temperature scaling, $T_{\rm
  mw}^{2/3}$. In all panels, the straight dotted line marks the slope
  $S\propto R^{0.95}$ which is the best--fit to the observed entropy
  profiles of Pratt \& Arnaud (2004) and Piffaretti et al. (2005). The
  vertical dotted lines are for $r_{200}$ and $r_{500}$ of the
  Cluster.}
\label{fi:redentr}
\end{figure*}

A further question we investigated is whether the different feedback
and pre--heating schemes are able to violate the self--similarity of
the entropy profiles as strongly as it is observed. For this purpose,
we plot in Figure~\ref{fi:redentr} the profiles of reduced entropy,
$S(r)/T_{\rm mw}(r)$, for the four simulated structures in different
radiative runs. Since in this analysis we do not aim for a detailed
comparison with observations, we here prefer to normalise the entropy
to the mass--weighted temperature, $T_{\rm mw}$, instead of using the
emission--weighted temperature. The latter has been shown to provide
an inaccurate measure for the spectroscopic temperature of the ICM
that is observationally inferred from X--ray spectra (e.g., Mathiesen
\& Evrard 2001; Mazzotta et al. 2004; Vikhlinin 2005). Although
cooling, star formation and SN feedback are likely to introduce
characteristic scales in the ICM thermodynamics, all the W1 runs have
remarkably self--similar entropy profiles, over the whole range of
resolved scales. However, for the W4 runs, the self--similarity is
clearly broken in the central regions, while it is largely preserved
at $R\magcir 0.5\,R_{\rm vir}$. Over the range of scales where
self--similarity is preserved, the slope of the entropy profiles is
always consistent with the $S\propto R^{0.94}$ behaviour found by
Pratt \& Arnaud (2004) and Piffaretti et al. (2005) from their
analysis of groups and clusters observed with XMM--Newton.

As expected, only the S100 entropy floor generates an appreciable
breaking of self--similarity in the outer regions of the simulated
structures. In the lower right panel of Figure~\ref{fi:redentr} we
show the entropy profiles rescaled according to $T_{\rm mw}^{2/3}$,
which is consistent with the scaling suggested by observational
data. Quite ironically, while this time we are able to reproduce the
scaling with temperature of the normalisation of the profiles, we lose
the agreement with the radial slope, which is significantly shallower
than the observed one.  We note here that an accurate derivation of
the ICM entropy profiles from observations requires properly resolving
both the gas density profiles and the temperature profiles. While
different authors generally agree on the radial run of gas density,
the temperature profiles are still a matter of vigorous debate. For
instance, both ASCA (e.g., Markevitch et al. 1998; Finoguenov et
al. 2001) and Beppo--SAX (e.g., De Grandi \& Molendi 2002)
consistently find decreasing profiles, at least for $R\, \magcir
0.2\,R_{180}$ (cf. also White 2000). However, Pratt \& Arnaud (2003,
2004) find profiles consistent with being isothermal, at least outside
the cooling regions. More recently, Piffaretti et al. (2005) and
Vikhlinin et al. (2004) find decreasing outer profiles from
XMM--Newton and Chandra data, respectively. While we do not intend to
enter this debate here, we note that shallower entropy profiles should
be obtained from data when the corresponding temperature profiles have
a negative outer gradient.

\section{Discussion and conclusions}
\label{s:disc}

In this paper, we have analysed hydrodynamical simulations of the
formation of one galaxy cluster and three galaxy groups, with the main
aim of studying the ability of different models of non--gravitational
heating to cause an amplification of entropy generation by accretion
shocks as a consequence of a transition from clumpy to smooth
accretion (e.g. Voit \& Ponman 2003). For this purpose, we have
carried out non--radiative simulations as well as simulations with
cooling and star formation, and combined those with two different
models for non--gravitational heating. The extra heating has been
supplied in two different forms: {\em (a)} by imposing an entropy
floor at $z_h=3$; {\em (b)} by including the effect of galactic winds
(Springel \& Hernquist 2003a) of different strengths in the runs with
star formation.

The main results of our analysis give answers to the three questions
posed in the Introduction, and can be summarised as follows.
\begin{description}
\item[(a)] Smoothing the accretion pattern by pre--heating in
  simulations with non--radiative physics does indeed amplify the
  entropy generation out to the radius where accretion shocks are
  taking place. As predicted by semi--analytic models (Voit et
  al. 2003), this amplification is more pronounced for lower--mass
  systems, because they accrete from smaller sub--halos whose gas
  content is more easily smoothed by extra heating.
\item[(b)] Radiative cooling reduces this amplification effect by a
  significant amount. Cooling has the effect of increasing the
  clumpiness of the accretion pattern, and thus acts against the
  smoothing induced by extra heating. We find that only our
  pre--heating model with an entropy floor $S_{\rm fl}=100$ keV cm$^2$
  generates a sizable entropy amplification out to the external
  regions, $R\,\magcir 0.5\,R_{\rm vir}$, of group--sized halos. 
\item[(c)] The heating from galactic winds is efficient in regulating
  star formation and in providing an increase of the ICM entropy in
  the central regions of groups. However, the winds have a negligible
  effect in smoothing the accretion pattern and thus do not trigger an
  appreciable entropy amplification effect. This result, which holds
  also for the strongest winds considered in our analysis, indicates
  that the role played by these winds tends to be fairly well
  localised around the star forming regions and therefore hardly
  affects the gas dynamics over the whole interior of dark matter
  halos.
\end{description}

Our results show that the temperature--entropy scaling provides
powerful constraints on the physical nature of the energy feedback
that affects the thermodynamic history of the diffuse
baryons. Smoothing the gas accretion pattern within halos of forming
galaxy clusters requires a rather non--local feedback, whose action
should not only be that of preventing gas overcooling. The
difficulties faced by our model for galactic winds suggest that we may
miss a proper description of physics which is able to distribute the
SN energy in the diffuse medium more efficiently (e.g. Kay et
al. 2003, 2004, where cold galactic gas was impulsively heated to high
temperature). Alternatively, this may be taken as an indication that
other astrophysical sources of energy feedback are required. In this
context, the most obvious candidate are probably AGN, whose effects
should be taken into account self--consistently in future cosmological
hydrodynamical simulations.

In this paper, we have not yet performed an in-depth comparison with
observational data because we preferred to first provide a general
interpretative framework for X--ray observations. However, a detailed
comparison with observations in future work is very promising for
extracting yet more information from the ICM
thermodynamics. Particularly the increasing amount of high quality
X--ray observations at the group scale (e.g. Finoguenov et al. 2004)
together with the ever more sophisticated simulation models give a
real hope that the intra--cluster baryons will remain an extremely
useful tracer and will eventually allow us to understand the nature of
feedback in groups and clusters.

\section*{Acknowledgements.} The simulations have been realized with
CPU time allocated at the ``Centro Interuniversitario del Nord-Est per
il Calcolo Elettronico'' (CINECA, Bologna), thanks to grants from INAF
and from the University of Trieste. This work has been partially
supported by the INFN Grant PD-51. We acknowledge useful discussions
with Klaus Dolag, Giuseppe Murante and Luca Tornatore.


\begin{thebibliography}{}

\bibitem[]{} Arnaud M., Evrard A.E., 1999, MNRAS, 305, 631
\bibitem[]{} Balogh M.L., Babul A., Patton D.R., 1999, MNRAS, 307, 463
\bibitem[]{} Bialek J.J., Evrard A.E., Mohr J.J., 2001, ApJ, 555, 597
\bibitem[]{} Borgani S., Governato F., Wadsley J., et al., 2001, ApJ,
  559, L71
\bibitem[]{} Borgani S., Governato F., Wadsley J., et al., 2002,
  MNRAS, 336, 409
\bibitem[]{} Borgani S., Murante G., Springel V., et al., 2004, MNRAS,
  348, 1078
\bibitem[]{} Bower R.G., 1997, MNRAS, 288, 355
\bibitem[]{} Bryan G.L., 2000, ApJ, 544, L1 
\bibitem[]{} Cavaliere A., Menci N., Tozzi P., 1998, ApJ, 501, 493
\bibitem[]{} De Grandi S., Molendi S., 2002, ApJ, 567, 163
\bibitem[]{} Dos Santos S., Dor{\'e} O., 2002, A\&A, 383, 450 
\bibitem[]{} Eke V.R., Navarro J., Frenk C.S., 1998, ApJ, 503, 569
\bibitem[]{} Evrard A.E., Henry J.P., 1991, ApJ, 383, 95
\bibitem[]{} Finoguenov A., Arnaud M., David L.P., 2001, ApJ, 555, 191 
\bibitem[]{} Finoguenov A., Borgani S., Tornatore L., B{\"o}hringer
  H., 2003, A\&A, 398, L35 
\bibitem[]{} Finoguenov A., Davis D.S., Zimer M., Mulchaey J.S., 2004,
  ApJ, submitted
\bibitem[]{} Finoguenov A., Jones C., B{\"o}hringer H., Ponman T.J., 
  2002, ApJ, 578, 74 
\bibitem[]{} Kaiser N., 1986, MNRAS, 222, 323
\bibitem[]{} Kaiser N., 1991, ApJ, 383, 104
\bibitem[]{} Kay S.T., Thomas P.A., Theuns T. 2003, MNRAS, 343, 608
\bibitem[]{} Kay S.T., Thomas P.A., Jenkins A., Pearce F.R., 2004,
  MNRAS, 355, 1091
\bibitem[]{} Lin Y.-T., Mohr, J.J., Stanford S.A., 2003, ApJ, 591, 749
\bibitem[]{} Mahdavi A., Finoguenov A., B\"ohringer H., Geller M.J.,
  Henry J.P., 2005, ApJ, 622, 187
\bibitem[]{} Markevitch M., 1998, ApJ, 504, 27
\bibitem[]{} Markevitch M., Forman W.~R., Sarazin C.~L., 
  Vikhlinin A., 1998, ApJ, 503, 77
\bibitem[]{} Mathiesen B.F., Evrard A.E., 2001, ApJ, 546, 100
\bibitem[]{} Mazzotta P., Rasia E., Moscardini L., Tormen G., 2004,
  MNRAS, 354, 10  
\bibitem[]{} Muanwong O., Thomas P.A., Kay S.T., Pearce F.R., 2002, 
  MNRAS, 336, 527
\bibitem[]{} Mushotzky R., Figueroa-Feliciano E., Loewenstein M.,
  Snowden S.L., 2003 (preprint astro--ph/0302267)
\bibitem[]{} Navarro J.F., Frenk C.S., White S.D.M., 1995, MNRAS, 275, 720 
\bibitem[]{} Navarro J.F., Frenk C.S., White S.D.M., 1997, ApJ, 490, 493 
\bibitem[]{} Piffaretti R., Jetzer Ph., Kaastra J.S., Tamura T., 2005,
  A\&A, 433, 101
\bibitem[]{} Ponman T.J., Cannon D.B., Navarro J.F., 1999, Nature,
397, 135
\bibitem[]{} Ponman T.J., Sanderson A.J.R., Finoguenov A., 2003,
  MNRAS, 343, 331
\bibitem[]{} Pratt G.W., Arnaud M., 2003, A\&A, A\&A, 408, 1
\bibitem[]{} Pratt G.W., Arnaud M., 2005, A\&A, 429, 791 
\bibitem[]{} Rosati P., Borgani S., Norman C., 2002, ARAA, 40, 539
\bibitem[]{} Sanderson A.J.R., Ponman T.J., Finoguenov A.,
  Lloyd-Davies E.J., Markevitch M., 2003, MNRAS, 340, 989
\bibitem[]{} Salpeter E.E., 1955, ApJ, 121, 161
\bibitem[]{} Springel V., Hernquist L., 2002, MNRAS, 333, 649
\bibitem[]{} Springel V., Hernquist L., 2003a, MNRAS, 339, 289 (SH03)
\bibitem[]{} Springel V., Hernquist L., 2003b, MNRAS, 339, 312
\bibitem[]{} Springel V., Yoshida N., White S.D.M., 2001, NewA, 6, 79
\bibitem[]{} Tornatore L., Borgani S., Springel V., Matteucci F.,
  Menci N., Murante G., 2003, MNRAS, 342, 1025
\bibitem[]{} Tozzi P., Norman C., 2001, 546, 63
\bibitem[]{} Vikhlinin A., 2005, ApJ, submitted (preprint
  astro--ph/0504098) 
\bibitem[]{} Vikhlinin A., Markevitch M., Murray S.S., Jones C.,
  Forman W., Van Speybroeck L., 2004, ApJ, submitted (preprint
  astro--ph/0412306)
\bibitem[]{} Voit, G.M. 2005, Rev. Mod. Phys., Jan. 2005 issue (preprint
  astro--ph/0410173) 
\bibitem[]{} Voit G.M., Balogh M.L., Bower R.G., Lacey C.G., Bryan
  G.L., 2003, ApJ, 593, 272
\bibitem[]{} Voit G.M., Bryan G.L., 2001, Nat, 414, 425
\bibitem[]{} Voit G.M., Bryan G.L., Balogh M.L., Bower R.G., 2002,
  ApJ, 576, 601
\bibitem[]{} Voit G.M., Ponman T.J., 2003, ApJ, 594, L75 
\bibitem[]{} White D.A., 2000, MNRAS, 312, 663 
\bibitem[]{} Wu X.-P., Xue Y.-J., 2002, ApJ, 569, 112

\end{thebibliography}
\end{document}